\documentclass[materials,review,submit,oneauthor,12pt,a4paper]{mdpi} 
\lastpage{x}
\doinum{10.3390/------}
\pubvolume{xx}
\pubyear{2011}
\history{Received: xx / Accepted: xx / Published: xx}

\usepackage{amssymb,amsmath}
\usepackage{graphicx}
\usepackage{color}

\Title{Buckling of Carbon Nanotubes: A State of the Art Review}

\Author{Hiroyuki Shima}

\address{Division of Applied Physics, Faculty of Engineering, Hokkaido University,
Kita-13, Nishi-8, Kita-ku, Sapporo, Hokkaido 060-8628 Japan}

\corres{Email: shima@eng.hokudai.ac.jp}

\abstract{The nonlinear mechanical response of carbon nanotubes, referred to as their ``buckling" behavior,
is a major topic in the nanotube research community.
Buckling means a deformation process in which a large strain beyond a threshold causes an 
\color{black}
abrupt change in the strain enregy vs. deformation profile.
\color{black}
Thus far, much effort has been devoted to analysis of the buckling of nanotubes
under various loading conditions: compression, bending, torsion, and their certain combinations.
Such extensive studies have been motivated by
(i) the 
\color{black}
structural 
\color{black}
resilience of nanotubes against buckling and
(ii) the substantial influence of buckling on their physical properties.
In this contribution, I review the dramatic progress in nanotube buckling research
during the past few years.}


\makeatletter
\AtBeginDocument{\@ifpackageloaded{natbib}{\ifNAT@numbers\if@filesw\immediate\write\@auxout{\string\global\string\NAT@numberstrue}\fi\fi}{}}
\makeatother
\begin{document}


\section{Introduction: Appeal of nanocarbon materials}\label{sec1}

Carbon is a rare substance that takes highly diverse morphology.
When carbon atoms form a three-dimensional structure,
their glittering beauty as diamonds is captivating.
When aligned in a two-dimensional plane,
they make up just black graphite and lose their sparkle.
In addition to these ``macro"-scopic carbon materials,
several ``nano"-carbon materials have been discovered
in the past few decades, opening up new horizons in material sciences.
It all began with the C$_{60}$ molecule (fullerene),
whose existence was predicted by Osawa \cite{OsawaKagaku1970} in 1970
and was discovered by Kroto {\it et al}. \cite{KrotoNature1985} in 1985.
Subsequent studies,
including those on carbon nanotubes by Iijima \cite{IijimaNature1991} 
in 1991\footnote{See 
\color{black}
Appendix
\color{black}
\ref{secapp}
for the delicate issue about who should be credited as the first person to discover carbon nanotubes.}
and
on graphene by Novoselov {\it et al}. \cite{NovoselovScience2004} in 2004, have had a tremendous impact
and driven developments in science and engineering around the turn of
the century.\footnote{Together with the three famous nanocarbons, ``nanodiamond" deserves attention
as the fourth member of nanocarbon materials, being classic yet novel.
Nanodiamond was initially synthesized by Volkov \cite{DanilenkoPSS2004} in 1963,
and recently, a wide variety of applications has been proposed;
see Refs.~\cite{OsawaPAC2008,Nanodiamond2010} for instance.}

Among the three types of nanocarbon materials, carbon nanotubes are
attracting the greatest attention in both industry and academia.
Research on carbon nanotubes has brought out two characteristics
not usually seen in other fields.
First and foremost is the sheer breadth of the research,
which encompasses physics, chemistry, materials science,
electronics, and life science.
The second characteristic is that basic research and applied research
are extremely close to each other.
A succession of phenomena of interest to scientists has been
discovered like a treasure chest, each leading to an innovative
application or development.
Nowadays, it is difficult even for professionals
in the nanotube research community
to understand the progress being made outside of their field of expertise.

One of the reasons why carbon nanotubes offer huge potential is the fact that
mechanical deformation causes considerable changes in electronic, optical, magnetic, and chemical properties. Thus, many studies on new technologies
to utilize the correlation between deformation and properties
are underway in various fields.
For example, studies of nanoscale devices based on the change in electrical
conductivity or optical response resulting from deformation are
one of the most popular trends in nanotechnology.
Another important reason for nanotube research diversity is
the concomitance of structural resilience and small weight, making realizable ultrahigh-strength materials
for utilization in super-high-rise
buildings and large aerospace equipment.
Furthermore, applications of these low-density substances
for aircraft and automobile parts will raise fuel efficiency and save energy,
as well as dramatically reduce exhaust gas emissions and environmental impact.

With this background in mind,
I shall review recent development in a selected area of 
nonlinear mechanical deformation,  the ``buckling" of carbon nanotubes.\footnote{A forthcoming
monograph \cite{ShimaPanPub},
which covers broad topics of carbon nanotube deformation, including buckling,
will be of great help to readers who are interested in the subject.}
Section \ref{sec2} provides a concise explanation of the terminology of buckling,
followed by a survey of different approaches used in  nanotube buckling investigations.
Section \ref{sec3} details  the two most interesting features observed
during nanotube buckling process, {\it i.e.,}  
\color{black}
the structural
\color{black}
resilience and sensitivity of nanotube properties
against buckling.
Sections \ref{sec4} to \ref{sec8} are the main part of this paper,
illustrating nanotube buckling under
axial compression (\S \ref{sec4}), radial compression (\S \ref{sec5}),
bending (\S \ref{sec6}, \ref{sec7}), and torsion (\S \ref{sec8}).
Section \ref{sec9} presents a universal scaling law
that describes different buckling modes of nanotubes in a unified manner.
\color{black}
The article is closed by Section 10 that describes 
several challenging problems whose solutions
may trigger innovation in the nanotube research community.
\color{black}
The list of references 
\color{black}
(over 190)
\color{black}
is fairly extensive,
although by no means all inclusive.
To avoid overlap with the existing excellent
reviews \cite{WatersCompSciTech2006,CYWangJNanosciNanotechnol2007,CMWangApplMechRev2010},
results reported within the past few years are featured
in words that nonspecialists can readily understand.



\section{Background of nanotube buckling research}\label{sec2}

\begin{figure}[ttt]
\centerline{
\includegraphics[width=6.0cm]{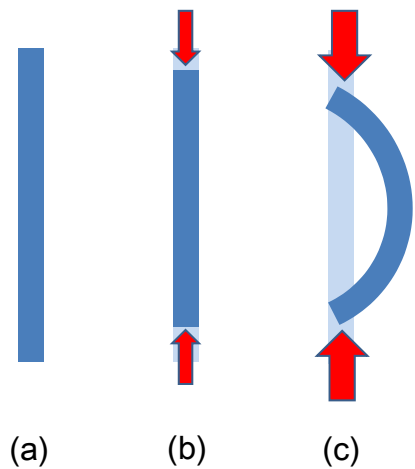}
}
\caption{Schematic diagram of buckling of an elastic beam
under axial compression:
(a) pristine beam; (b) axial compression for a small load;
(c) buckling observed beyond a critical load.
}
\label{axialbucklingFig}
\vspace*{60pt}\end{figure}

The term ``buckling'' means a deformation process in which
a structure subjected to high stress undergoes a sudden 
\color{black}
change in morphology at a critical load 
\color{black}
\cite{Brush}.
A typical example of buckling may be observed
when pressing  opposite edges of a  long, thin elastic beam toward one another;
see Fig.~\ref{axialbucklingFig}.
For small loads, the beam is compressed in the axial direction
while keeping its linear shape [Fig.~\ref{axialbucklingFig}(b)],
and the strain energy is proportional to the square of the axial displacement.
Beyond a certain critical load, however,
it suddenly bends archwise [Fig.~\ref{axialbucklingFig}(c)]
and the relation between the strain energy and displacements
deviates significantly from the square law.
Besides axial compression,
bending and torsion give rise to
buckling behaviors of elastic beams,
where the buckled patterns strongly depend on
 geometric and material parameters.
More interesting is the elastic buckling of structural pipe-in-pipe cross sections
under hydrostatic pressure \cite{SatoJMST2007,SatoStrucEngMech2008};
pipe-in-pipe ({\it i.e.,} a pipe inserted inside another pipe) applications
are promising for offshore oil and gas production systems
in civil engineering.

The above argument on macroscopic elastic objects
encourages to explore what buckled patterns
are obtained in carbon nanotubes.
Owing to their nanometric scales,
similarities and differences in buckled patterns compared with macroscopic counterparts
should not be trivial at all.
This complexity has motivated tremendous efforts
toward the buckling analysis of carbon nanotubes under diverse loading
conditions:
axial compression \cite{YakobsonPRL1996,RuJAP2000,RuJMPS2001,RuJAP2001,BNiPRL2002,BuehlerASME2004,
WatersAPL2004,PantanoASME2004,WatersAPL2005,SearsPRB2006,YYZhangJAP2008},
radial compression \cite{VenkateswaranPRB1999,DYSunPRB2004,
J_Tang2000,Peters2000,Sharma2001,Rols2001,Reich2002,PantanoJMPS2004,
CYWangJNanosciNanotechnol2003,
ElliottPRL2004,Tangney2005,Gadagkar2006,XWangPRB2006,SZhangPRB2006,HasegawaPRB2006,
YHuangPRB2006,XYangAPL2006,Chrisofilos2007,ImtaniPRB2007,
Peng2008,Giusca2008,Wu2008,Jeong2008,ZXuSmall2008,SHYangJPD2009,
Kuang2009,Lu2009,ShimaNTN2008,ShimaPSSA2009,ShimaPRB2010,ShimaCMS2011,XHuangNanoResLett2011,
HYWangAPL2006,BarbozaPRL2009,YHYangAPL2011},
bending \cite{IijimaJCP1996,FalvoNature1997,PoncharalScience1999,DuanNanoLett2007,
ShibutaniModelSimulMaterSciEng2004,KutanaPRL2006,YangModelSimulMaterSciEng2006,XWangPRB2006,QWangAPL2007},
torsion \cite{BWJeongAPL2007,HKYangComposStric2007,YYZhangJPCM2008,QWangCarbon2008,QWangCarbon2009,BWJeongCarbon2010},
and their certain combinations \cite{XWangNTN2006,YJLuJPD2006,XWangIntJSolidStruc2007,
CLZhangPRB2007,BWJeongAPL2008}.
Such  extensive studies have been driven primarily by the following two facts.
One is 
\color{black}
the excellent geometric reversibility
\color{black}
of nanotubes against mechanical deformation; 
that is, their 
\color{black}
cylindrical
\color{black}
shapes are reversible upon unloading
\color{black}
without permanent damage to the atomic structure. In addition,
\color{black}
carbon nanotubes exhibit high fatigue resistance;
therefore, they are the promising medium for the mechanical energy storage with extremely
high energy density \cite{RZhangAdvMater2011}.
The other fact is the substantial influence of buckling on their physical properties.
It was recently shown that, just as one example,
carbon nanotubes undergoing an axial buckling instability
have  potential utility as a single-electron transistor \cite{WeickPRB2010,NaieniJPD2011,WeickPRB2011} 
and can play a crucial role in developing nanoelectromechanical systems.

Microscopy measurements are powerful means of examining
the nonlinear response of nanotubes against external loading.
For instance, atomic force microscopy (AFM) was utilized to
reveal the force--distance curve of nanotubes while buckling \cite{YapNanoLett2007}.
More direct characterizations of nanotube buckling were obtained by
{\it in situ} transmission electron microscopy (TEM) \cite{KuzumakiJJAP2006,MisraPRB2007,JZhaoCarbon2011},
as partly demonstrated in Fig.~\ref{KuzumakiJJAP2006Fig2} (see \S \ref{sec4sub2}).
However, the experimental investigation of nanotube buckling
remains a challenge because of difficulties in manipulation
at the nanometric scale.
This is a reason why both theoretical and
numerical approaches have played an important role in exploring
the buckling behavior of nanotubes.
In theoretical studies, carbon nanotubes are commonly treated as beams
or thin-shell tubes with certain wall thickness and elastic constants
\cite{RuPRB2000,CYWangIJSS2003,HSShenIJSS2004,RafiiTabarPhysRep2004,XQHeJMPS2005,
SilvestreIJSS2008,KulathungaJPCM2009,SilvestreCompStr2011,
KudinPRB2001,PantanoJMPS2004,YHuangPRB2006,ChandrasekerCMS2007}.
Such  continuum approximations are less computationally expensive than atomistic approaches;
moreover, the obtained formulations can be relatively simple in general.
It is noteworthy that, by substituting appropriate values
into elastic constants,
continuum-mechanics approaches provide a unified framework \cite{SilvestreIJSS2008}
that accounts for the critical buckling strains and buckled morphologies
under various loading conditions covering compression,
bending, torsion, etc.


\section{Resilience and sensitivity to buckling}\label{sec3}

Special emphasis should be placed on the fact that
carbon nanotubes exhibit many intriguing postbuckling morphologies:
Radial corrugations (see \S \ref{sec5sub2})
and surface rippling (\S \ref{sec9}) are typical examples.
One of the most outstanding features of postbuckled nanotubes
is their 
\color{black}
geometric reversibility upon unloading.
\color{black}
Indeed, experiments have shown that
the buckling deformation can be completely recovered
when the load is
removed \cite{IijimaJCP1996,FalvoNature1997,KnechtelAPL1998,LouriePRL1998,BowerAPL1999,PoncharalScience1999,TomblerNature2000}.
The marked 
\color{black}
structural
\color{black}
resilience is primary because of
(i) the large in-plane rigidity of graphene sheets
rather than low bending rigidity \cite{CLeeScience2008}
and (ii) the intrinsic hollow geometry with extremely large aspect ratio
that carbon nanotubes exhibit.
It was suggested that
the resilience makes it possible to use the nanotubes as a
gas  \cite{ZQZhangPRB2008,QWangNanoLett2009,QWangCarbon2009}
or water pipeline \cite{WHDuanASCNano2010} whose permeability can be tuned by mechanical deformation.

Apart from the 
\color{black}
structural
\color{black}
resilience, the sensitivity of carbon nanotube properties to buckling
is worthy of attention.
In fact, the breakdown of the structural symmetry resulting from the buckling triggers sudden
changes in physical and mechanical properties of nanotubes,
including
 thermal conductivity reduction \cite{ZXuNTN2009,XLiPRB2010,ZHuangJAP2011},
a radial breathing-mode frequency shift \cite{WYangPRB2008},
the emergence of interwall sp$^3$ bondings \cite{CTangJAP2010},
and electromechanical responses under bending \cite{RochefortPRB1999,BozovicAPL2001,FarajianPRB2003}
and torsion \cite{HallNatureNTN2007,DBZhangPRB2010}, to name a few.
In addition, the buckling-induced reduction in nanotube stiffness not only
impairs the ability of nanotubes to sustain external loadings as reinforced fibers
in nanocomposites \cite{LouriePRL1998,BowerAPL1999}
but also gives rise to large uncertainties
in the vibration behavior of nanotubes as nanoscale
resonators \cite{PoncharalScience1999,PetersPRB2000,HallNanoLett2008}.
These buckling--property relations can significantly influence the performance of nanotubes
as structural or functional elements,
thus implying the need of a huge amount of effort 
that has been made for the study
of nanotube buckling.

\section{Axial compression buckling}\label{sec4}

\subsection{Shell buckling or column buckling?}\label{sec4sub1}

\begin{figure}[ttt]
\centerline{
\includegraphics[width=7.0cm]{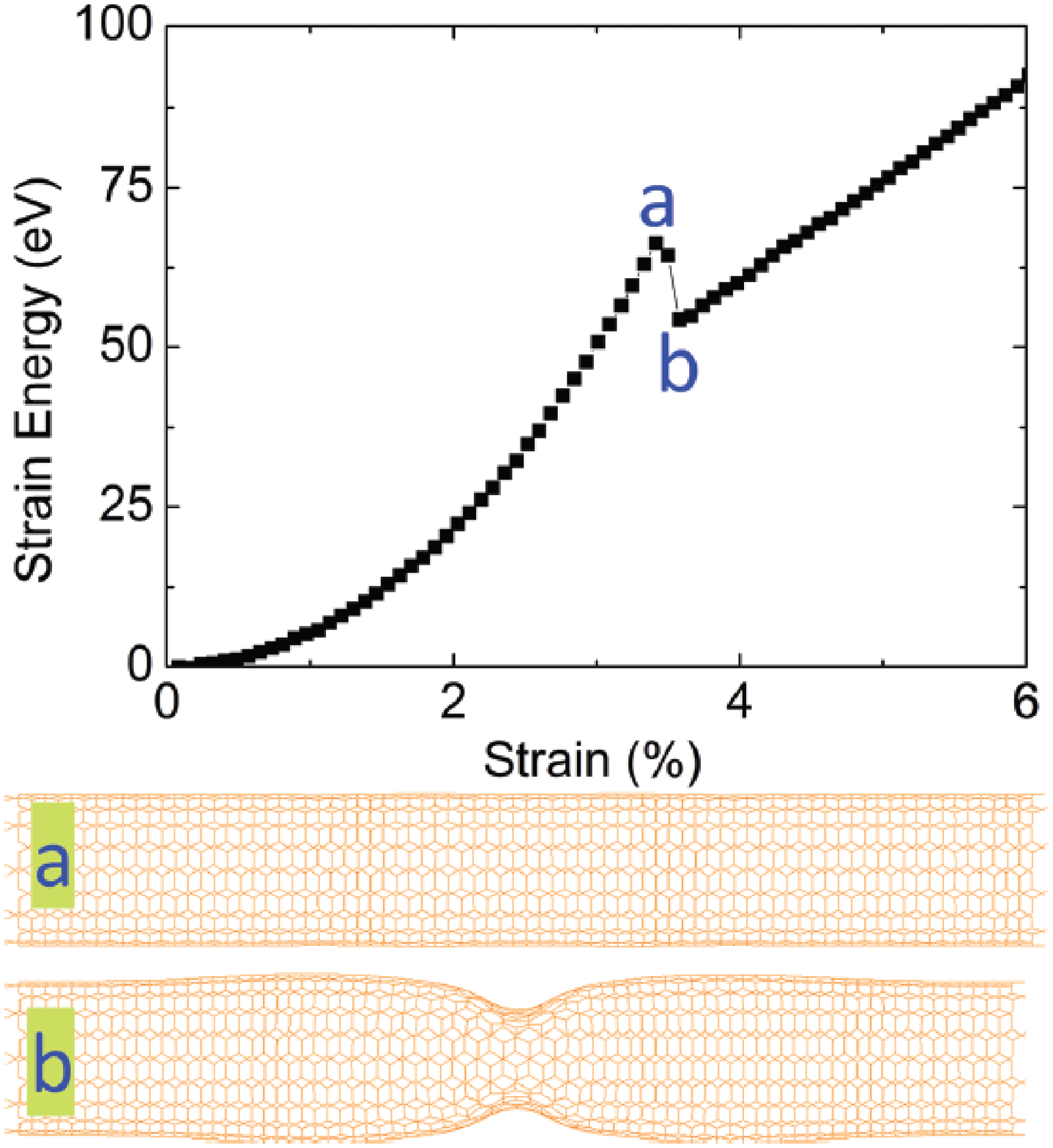}
\includegraphics[width=7.0cm]{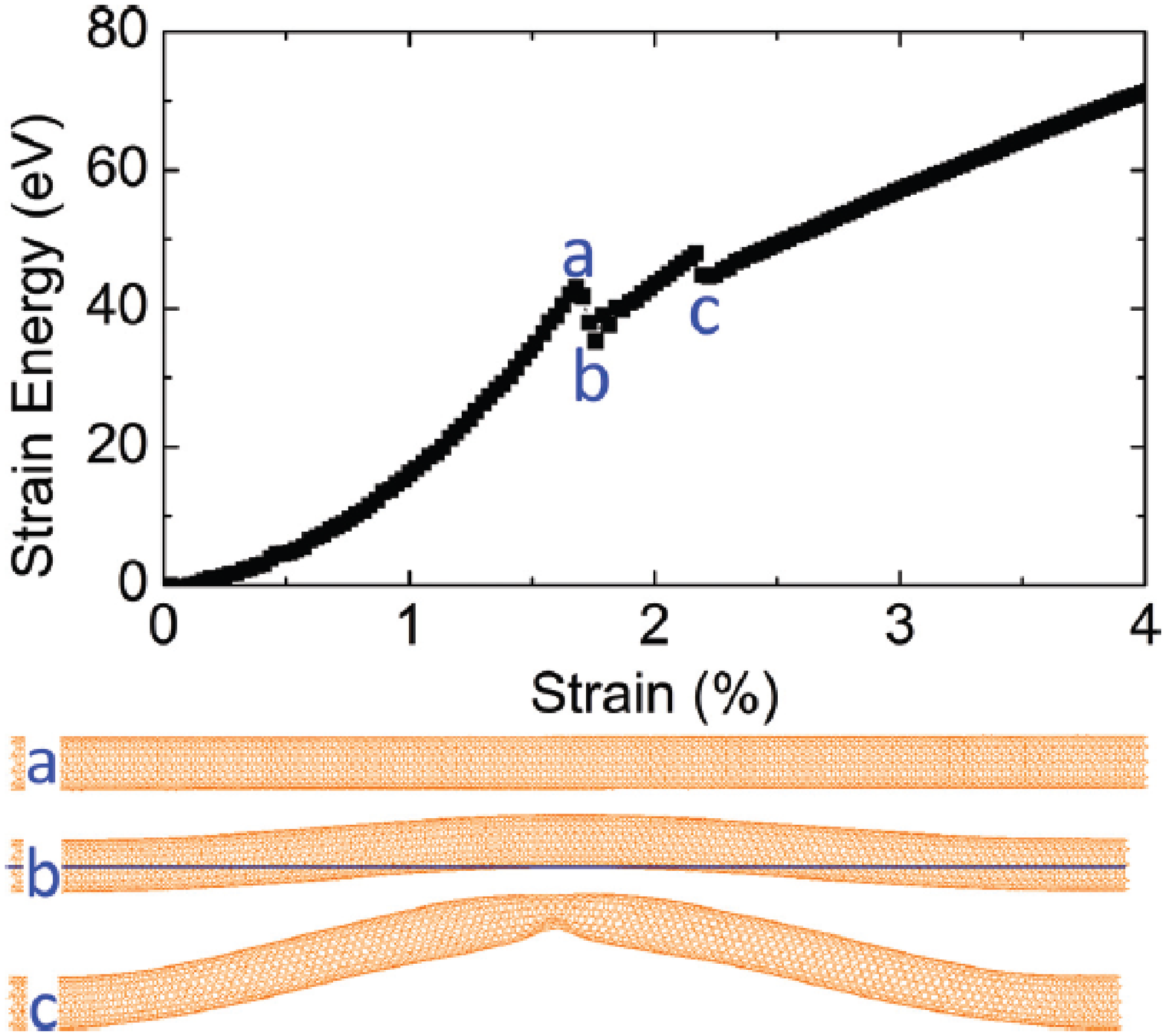}
}
\caption{
\color{black}
Axially buckled SWNT pattern deduced from molecular dynamics simulations.
\color{black}
[Left] Upper panel: Energy-strain curve of a (10,10)
SWNT with a length of 9.6 nm under axial compression.
Lower panel: Typical tube geometry (a) before and (b) after buckling,
respectively.
[Right]  Upper panel: Curve of a compressed (10,10) SWNT 
with 29.5 nm length.
Lower panel: Snapshots
of the tube (a) before buckling, (b) after column buckling,
and (c) undergoing shell buckling.
Reprinted from Ref.~\cite{FelicianoJAP2011}.
}
\label{Fig_FelicianoJAP2011}
\vspace*{60pt}\end{figure}

Buckled patterns of single-walled nanotubes (SWNTs) under axial compression  depend on
their aspect ratio \cite{YakobsonPRL1996,RuJAP2001,PantanoJMPS2004},
which equals the ratio of length to diameter of nanotubes.
Roughly, a thick and short SWNT ({\it i.e.,} with small aspect ratio)
undergoes shell buckling while keeping a straight cylindrical axis,
whereas a thin and long one tends to exhibit a series of shell and column (or Euler)
buckling.

The shell buckling process is depicted in the left panel of 
Fig.~\ref{Fig_FelicianoJAP2011} \cite{FelicianoJAP2011},
where a (10,10) SWNT with a  length of 9.6 nm and an aspect ratio of $\sim$7
was chosen.\footnote{
\color{black}
Recently, shell buckling behavior of short nanotubes with aspect ratior $\sim$1 or less
was considered, showing the significant dependence of the buckling strain
on the nanotube length \cite{KorayemProcEng2011}. 
\color{black}
}
It is seen that the strain energy increases quadratically with strain
at the early prebuckling stage. At a critical strain of 3.5\%, a
sudden drop in energy is observed,\footnote{The critical strain of shell buckling 
is inversely proportional to the tube diameter \cite{YYZhangNTN2009}.}
corresponding to the occurrence of shell buckling.
During the postbuckling stage, the strain energy exhibits a linear relationship with strain.
\color{black}
The linear growth in energy is understood by
the primary role of the change in carbon-carbon (C-C) bond angles,
rather than that of the bond length variation,
in determining the energy-strain relation after buckling.
Detailed analyses in Ref.~\cite{FelicianoJAP2011} showed that within the post-buckling regime,
the variation in bond angles between neighboring C-C bonds becomes significant 
while C-C bond lengths show less variation;
this results in an almost constant axial stress, as deduced from Fig.~\ref{Fig_FelicianoJAP2011}.
\color{black}

With  increasing aspect ratio, the buckling mode 
switches to  a column buckling mode owing to the increased flexibility of the tube.
Column buckling is seen in the right panel of Fig. \ref{Fig_FelicianoJAP2011}
for a much longer (10,10)
SWNT (29.5 nm in length with an aspect ratio of $\sim$22).
A sudden drop in strain energy occurs at a strain level of 1.6\%,
beyond which the center is displaced in a transverse direction
away from its original cylindrical axis.
When  the already column-buckled SWNT is further compressed, the structure curls further and a second
drop in strain energy is observed at a strain level of
2.2\%.
The second drop corresponds to the onset of another
buckling, which is responsible for releasing the excess strain
energy. The tube geometry at this point indicates that the
SWNT undergoes  shell buckling. With further axial compression,
a linear relationship is observed between energy and strain.
This result indicates a column- to shell-buckling transition of SWNTs 
with large aspect ratio.\footnote{Similar sequential transitions can be 
observed when a SWNT is bent \cite{KutanaPRL2006}.
See \S \ref{sec6sub3} for details.}

\begin{figure}[ttt]
\centerline{
\includegraphics[width=8.0cm]{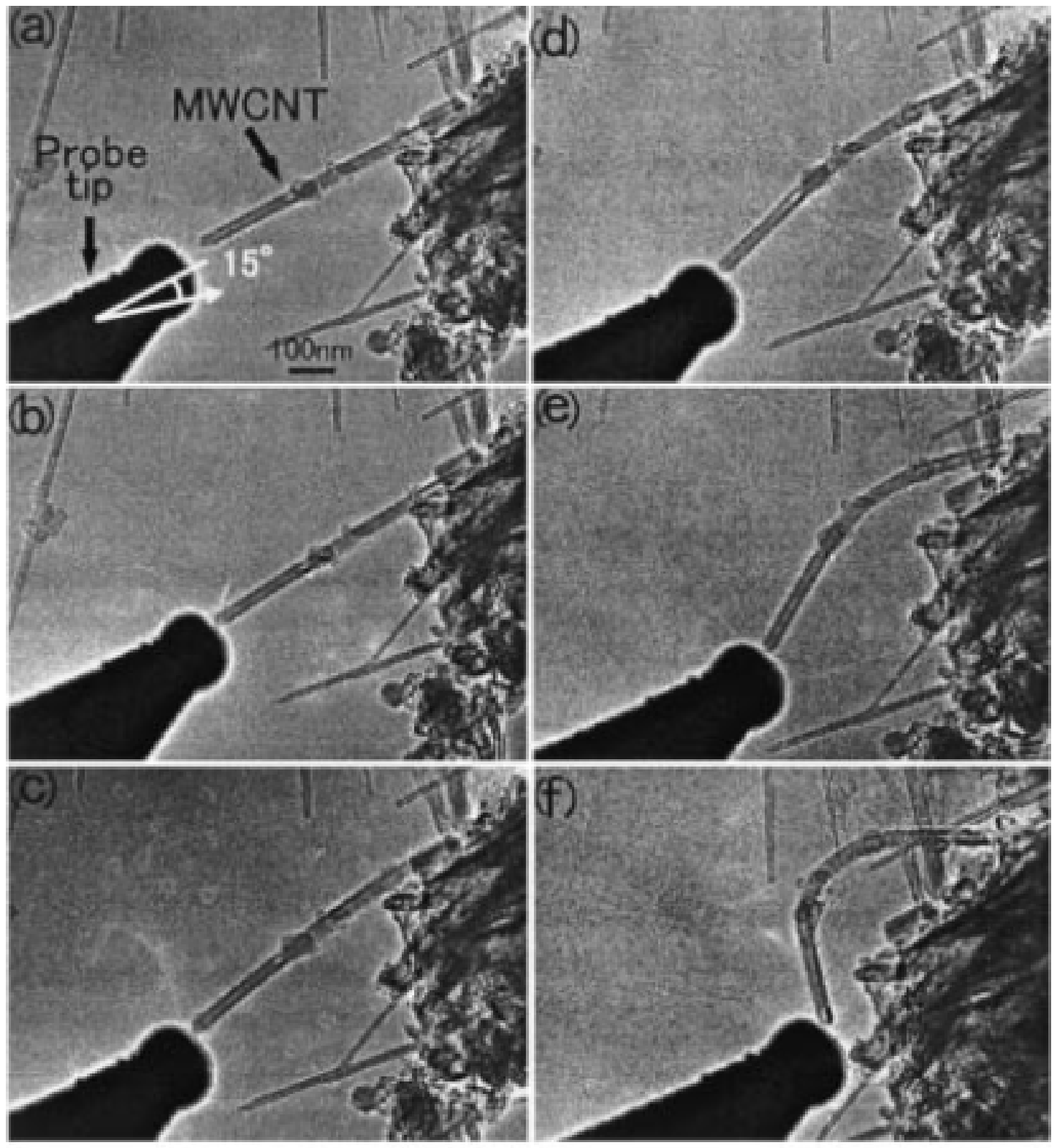}
\includegraphics[width=7.8cm]{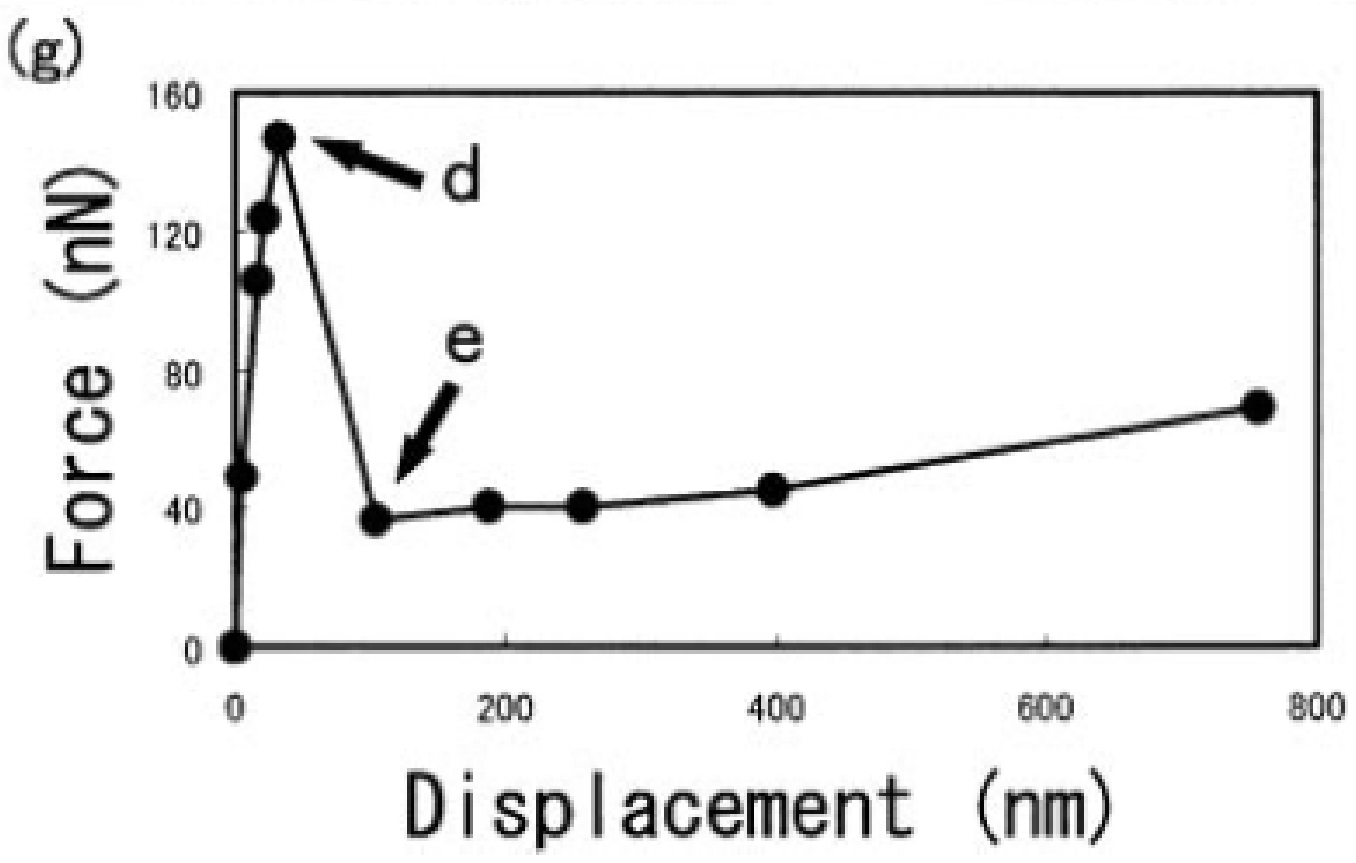}
}
\caption{(a)--(f) Series of TEM images of deformation processes for MWNTs
initiated by applying compressive force in the sample direction.
(g) Force--displacement diagram. The points indicated by arrows
correspond to the TEM images in (d) and (e).
Reprinted from Ref.~\cite{KuzumakiJJAP2006}.
}
\label{KuzumakiJJAP2006Fig2}
\vspace*{60pt}\end{figure}

It is important to note that the initial buckling modes,
corresponding to the first drop in energy-strain curve,
are different between large- and small-aspect-ratio SWNTs.
This fact necessitates an examination of the validity of continuum-mechanics models
for the buckling of SWNTs.
Careful assessments of the continuum approximations have been
reported \cite{SilvestreIJSS2008,KulathungaJPCM2009,YYZhangNTN2009,SilvestreCompStr2011,ArashCMS2012},
indicating the need to properly use  different models depending on the aspect ratio.
As to their consistency with atomistic simulation results,
readers can refer to Ref.~\cite{YYZhangNTN2009} in which
a list of critical strain data under different conditions
is detailed.

\vspace*{1cm}

\subsection{Force--displacement curve}\label{sec4sub2}

We now turn to experimental facts \cite{XHuangNanoRes2010}.
Because of the difficulty in  sample preparation and manipulation,
only a few attempts have been made to perform axial buckling
measurements \cite{KuzumakiJJAP2006,MisraPRB2007,JZhaoCarbon2011}.
\color{black}
In particular, experimental realization of shell buckling under compression
has largely behind, though its signature has been obtained via nanoindentation
\cite{GuduruExpMech2007}.
Hence in the following discussion, we focus our attention to the column buckling measurements.
\color{black}

The pioneering work \cite{KuzumakiJJAP2006} is presented in Fig.~\ref{KuzumakiJJAP2006Fig2};
The TEM images of (a)--(f) clarify a series of
deformation processes for  multiwalled nanotubes (MWNTs) initiated by applying a compressive force
in the nearly axial direction.\footnote{The actual direction of the applied force
deviates from the exact axial one by approximately 15$^\circ$,
as indicated in Fig.~\ref{KuzumakiJJAP2006Fig2}(a).}
Figure \ref{KuzumakiJJAP2006Fig2}(g) shows the corresponding force--displacement diagram.
The force at the initial stage is almost proportional to the displacement
[left to the point (d)],
indicating the elastic region, followed by an abrupt decrease at (e).
The two points indicated in Fig. 3(g) correspond to the TEM images in  panels (d) and (e), 
respectively.
\color{black}
In Fig.~\ref{KuzumakiJJAP2006Fig2}(g), the curve right to the point (e) maintains
a slight upward slope.
The reason for this post-buckling strength may be due to sequential emergence
of different buckling patterns with increasing the displacement.
\color{black}

\begin{figure}[ttt]
\centerline{
\includegraphics[width=12.0cm]{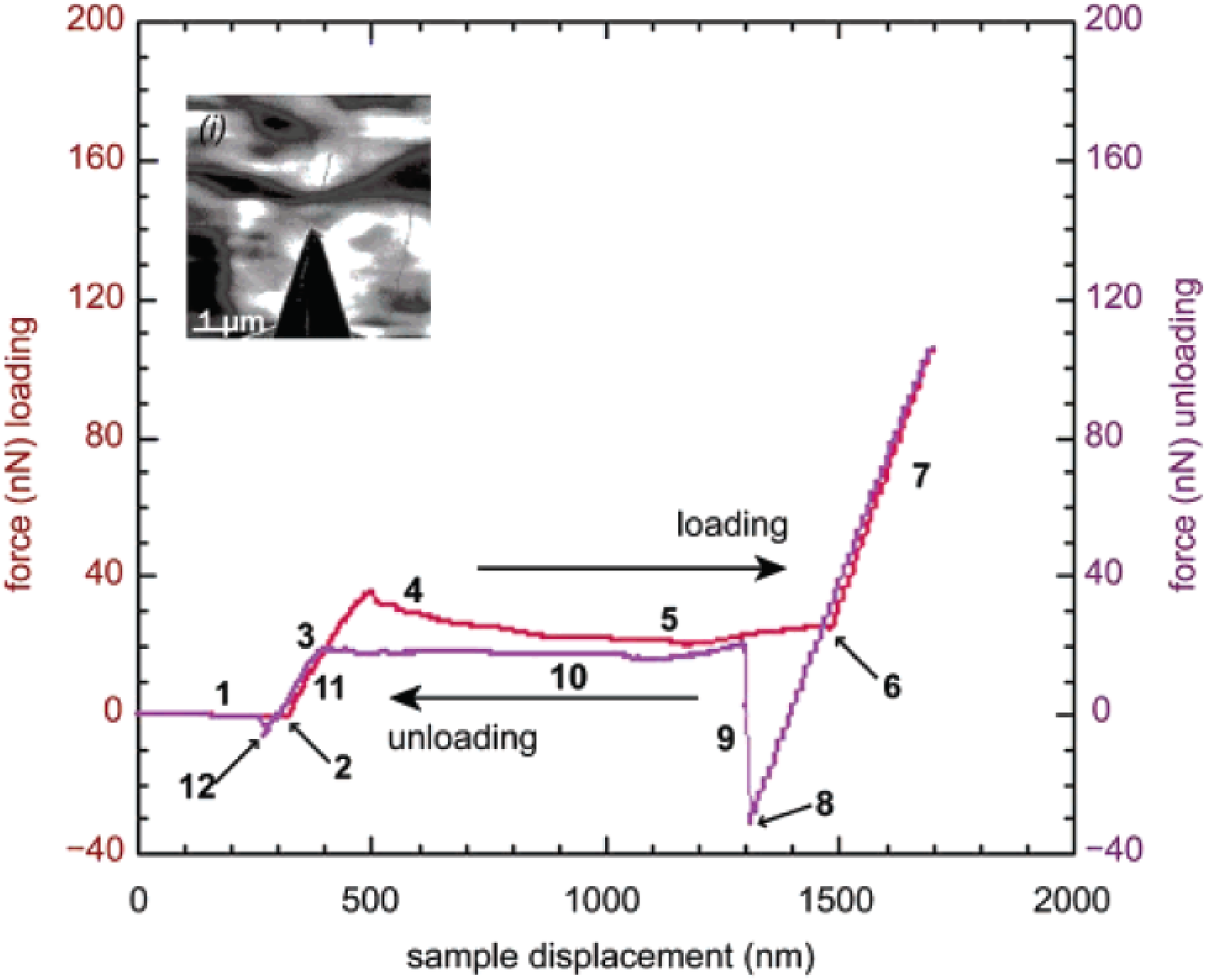}
}
\vspace*{6pt}
\centerline{
\includegraphics[width=11.5cm]{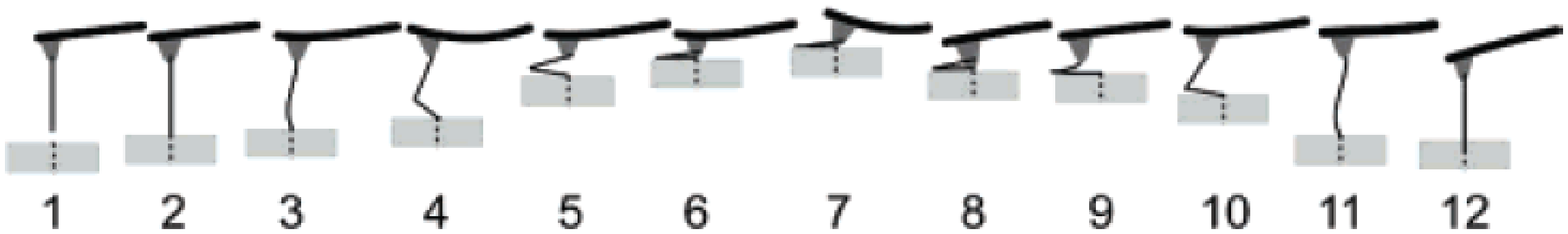}
}
\caption{
[Top]
Force--displacement curve of an MWNT with an aspect ratio of $\sim$80 under cyclic axial loading.
This inset shows a microscopy image.
[Bottom]
Schematic of the change in the MWNT configuration during the buckling process.
The labels correspond to those indicated in the force--displacement curve.
Reprinted from Ref.~\cite{YapNanoLett2007}.
}
\label{YapNanoLett2007Fig}
\vspace*{60pt}\end{figure}

A more sweeping measurement on the nanotube resilience was performed for the MWNT
with a higher aspect ratio ($\sim$80).
Figure \ref{YapNanoLett2007Fig} shows the resulting force--displacement curve
and graphical illustration of the buckling process \cite{YapNanoLett2007}.
An important observation is a negative stiffness region (labeled by ``4" in the plot)
that begins abruptly.
The sharp drop in force with increasing axial strain,
observed at the boundary of  regions (3) and (4),
is attributed to the kinking of the MWNT as depicted in the lower panel.
After the kinking takes place, the system is mechanically 
\color{black}
instable; this behavior is consistent with the mechanics of kinking described 
in Refs.~\cite{YakobsonPRL1996,KuzumakiJJAP2006}.
\color{black}
The instability seen in region (4) is reproducible through cyclic compression,
which opens up the possibility of harnessing the resilient mechanical properties of MWNTs
for novel composites.\footnote{The instability at (8) and the sharp rise at (9) 
during unloading stem from
the tip pulled out of contact from the sample, while
the nanotube end remains in contact.}

\section{Radial compression buckling}\label{sec5}

\subsection{Uniaxial collapse of SWNTs}\label{sec5sub1}

Radial pressure can yield a distinct class of buckling,
reflecting the high flexibility of graphene sheets in the 
\color{black}
normal
\color{black}
direction.
In fact, radial stiffness of an isolated carbon nanotube is much less than
axial stiffness \cite{PalaciPRL2005}, which results in an elastic deformation of
the cross section on applying
hydrostatic pressure \cite{VenkateswaranPRB1999,J_Tang2000,Peters2000,Sharma2001,Rols2001,
Reich2002,DYSunPRB2004,ElliottPRL2004,Tangney2005,Gadagkar2006,SZhangPRB2006,HasegawaPRB2006,XYangAPL2006,
Chrisofilos2007}
or indentation \cite{MajidJAP2008,BarbozaPRL2009,YHYangAPL2011}.
Experimental and theoretical studies, focused on SWNTs and their bundles,
revealed flattening and polygonalization in their cross section
under pressures on the order of a few gigapascals  \cite{VenkateswaranPRB1999,Sharma2001}.
Nevertheless, existing results are rather scattered, and we are far away from a unified understanding;
for example, the radial stiffnesses of nanotubes estimated thus far vary by up to three orders of
magnitude \cite{Reich2002,ElliottPRL2004,DYSunPRB2004,PalaciPRL2005,HYWangAPL2006,HasegawaPRB2006,
BarbozaPRL2009,YHYangAPL2011}.

\begin{figure}[ttt]
\centerline{
\includegraphics[width=9.5cm]{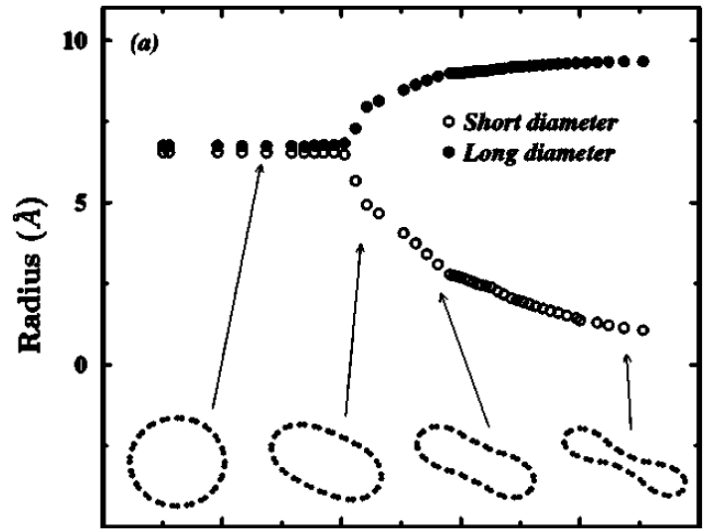}
}
\caption{Long and short diameters of a (10,10) SWNT as a function
of applied hydrostatic pressure. The shape of the cross section at
some selected pressures is plotted at the bottom of the figure.
Reprinted from Ref.~\cite{DYSunPRB2004}.
}
\label{DYSunPRB2004Fig2}
\vspace*{60pt}\end{figure}

The overall scenario of SWNT deformation under hydrostatic pressure
is summarized in Fig.~\ref{DYSunPRB2004Fig2} \cite{DYSunPRB2004}.
With increasing pressure, cross sections of SWNTs deform continuously from
circular to elliptical, and finally to peanut-like configurations
\cite{DYSunPRB2004,SZhangPRB2006,WLuPRB2011}.\footnote{For larger radius SWNTs,
the peanut-like deformed structure can be
transformed to dumbbell-like configurations by van der
Waals (vdW) attractions between the opposite walls of nanotubes.
The latter structure is energetically stable even when the applied
force is unloaded.}
The radial deformation of carbon nanotubes strongly affects their physical and structural properties.
For instance, it may cause semiconductor--metal transition \cite{BarbozaPRL2008,GiuscaNanoLett2008},
optical response change \cite{ThirunavukkuarasuPRB2010},
and magnetic moment quenching \cite{DinizPRB2010} in nanotubes.
From a structural perspective,
the radial collapse can give rise to interwall sp$^3$ bonding between
adjacent concentric walls \cite{FonsecaPRB2010,SakuraiPhysicaE2011},
which may increase nanotube stiffness and therefore be effective for high-strength
reinforced composites \cite{ZHXiaAPL2007,ByrnePRL2009,FilleterAdvMater2011,YYZhangJAP2011}.

\begin{figure}[ttt]
\centerline{
\includegraphics[width=10.5cm]{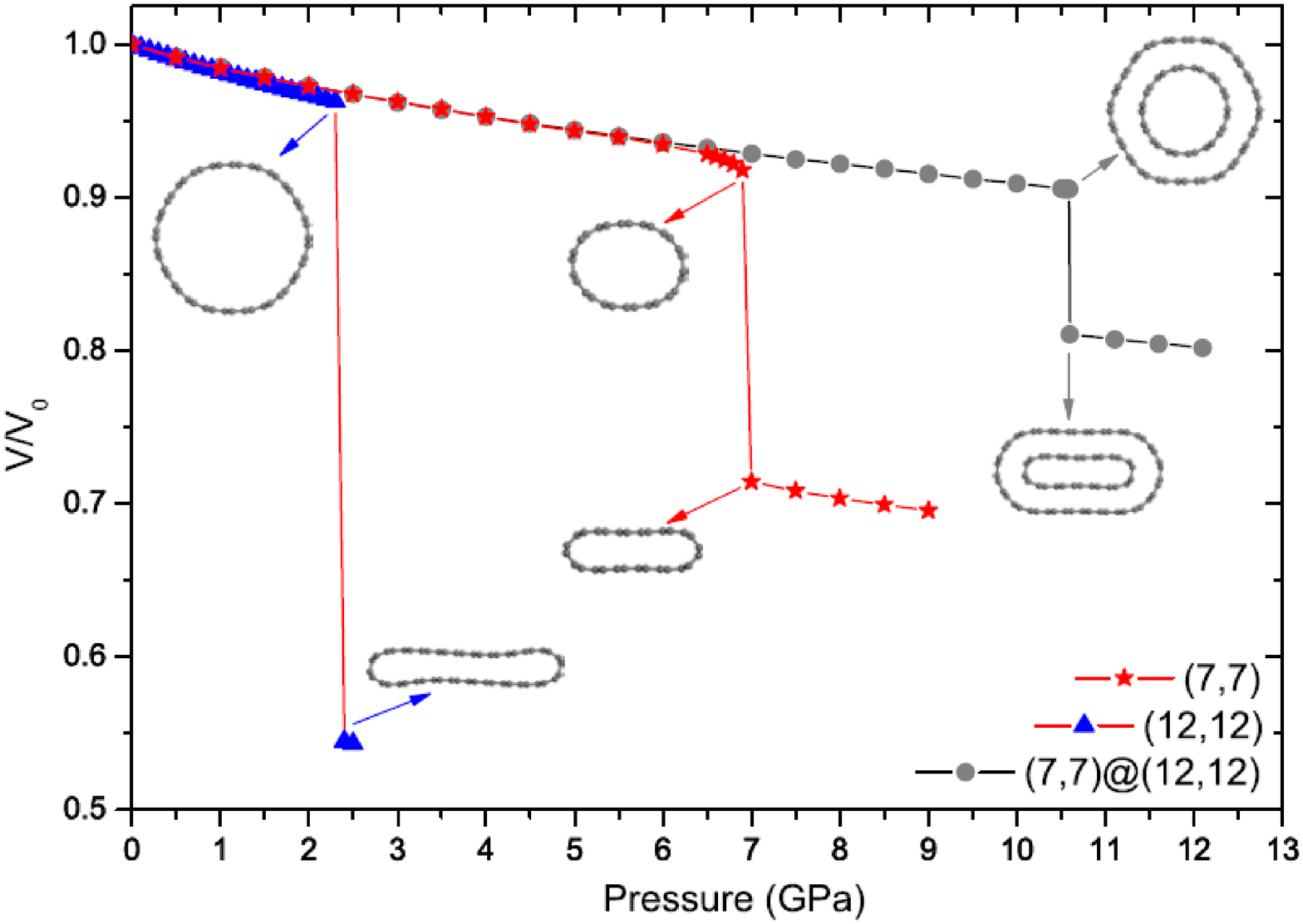}
}
\caption{Change in the relative volume of the (7,7)@(12,12)
DWNT bundle and the corresponding SWNT bundles as a
function of hydrostatic pressure.
Reprinted from Ref.~\cite{XYangEPL2008}.
}
\label{XYangEPL2008Fig}
\vspace*{60pt}\end{figure}

A bundle of nanotubes ({\it i.e.,} an ensemble of many nanotubes arranged parallel to each other)
can exhibit similar radial collapse patterns to those of an isolated nanotube 
under hydrostatic pressure.
Figure \ref{XYangEPL2008Fig} shows \cite{XYangEPL2008}
the volume change of a bundle of (7,7) SWNTs and
a bundle of (12,12) SWNTs as a function of the applied hydrostatic pressure;
the data for a bundle of (7,7)@(12,12) double-walled nanotubes (DWNTs) is also shown in the same plot.
The (12,12) SWNT bundle, for instance, collapses spontaneously at a critical pressure of 2.4 GPa,
across which the cross section transforms into a peanut-like shape.
Two other bundles provide higher critical pressures, as follows from the plot.
An interesting observation is that
the transition pressure of the (7,7) tube,
which is nearly 7.0 GPa when the tube is isolated,
becomes higher than 10.5 GPa when it is surrounded by the (12,12) tube.
This means that the outer tube acts as a
``protection shield" and the inner tube supports the outer
one and increases its structural stability;
this interpretation is consistent with the prior optical spectroscopic measurement
\cite{ArvanitidisPRB2005}.
This effect, however, is weakened as the tube radius increases
owing to the decreasing radial stiffness of SWNTs.

\vspace*{12pt}

\subsection{Radial corrugation of MWNTs}\label{sec5sub2}

In contrast to the intensive studies on SWNTs (and DWNTs),
radial deformation of MWNTs remains relatively unexplored.
Intuitively, the multilayered structure of
MWNTs is expected to enhance the radial stiffness relative to a single-walled counterpart.
However, when the number of concentric walls is much greater than unity, outside walls have
large diameters, so external pressure may lead to a mechanical instability
in the outer walls.
This local instability triggers a novel cross-sectional deformation, called
radial corrugation \cite{ShimaNTN2008}, of MWNTs
under hydrostatic pressure.

\begin{figure}[ttt]
\centerline{
\includegraphics[width=7.5cm]{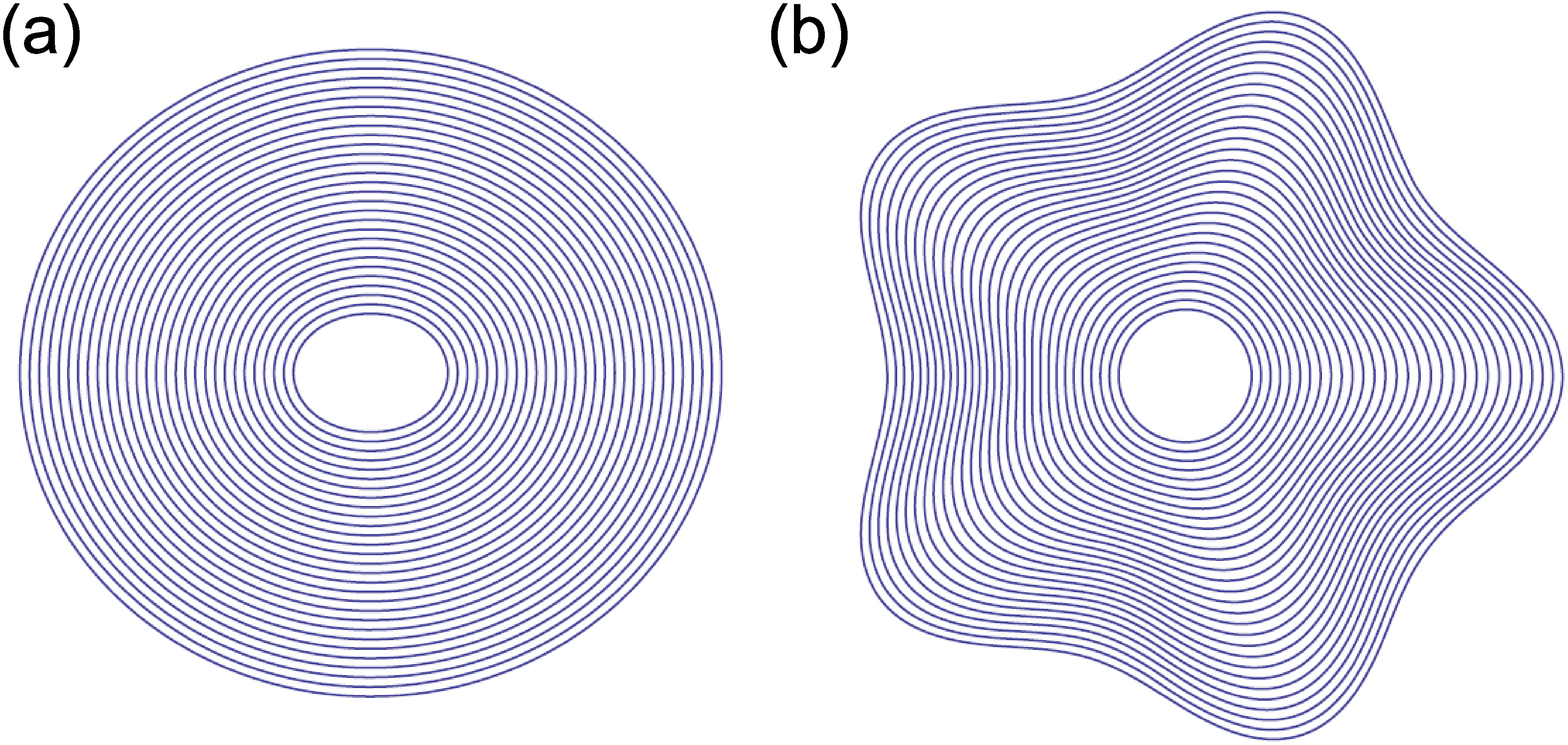}
}
\vspace*{6pt}
\centerline{
\includegraphics[width=6.5cm]{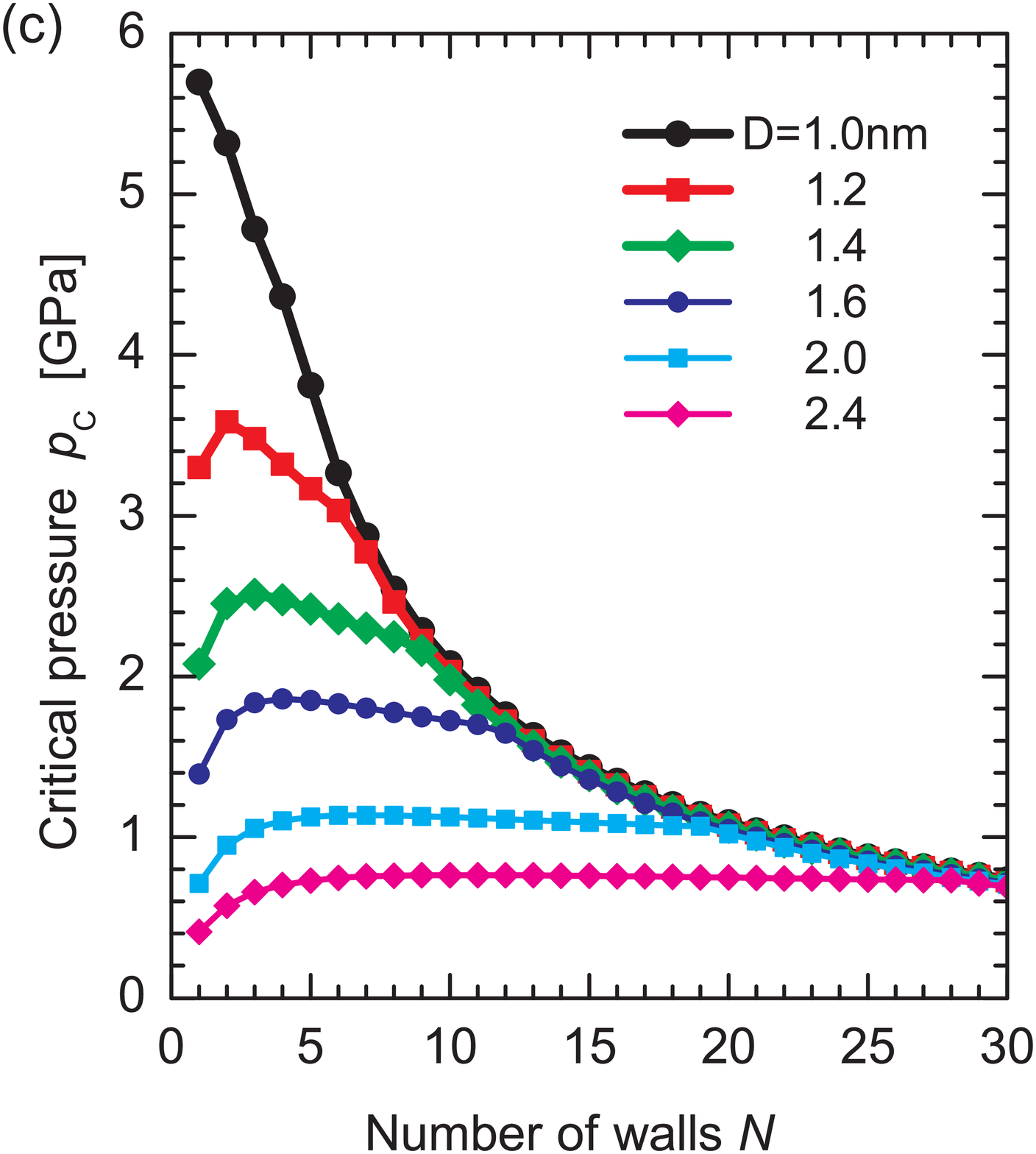}
\includegraphics[width=6.5cm]{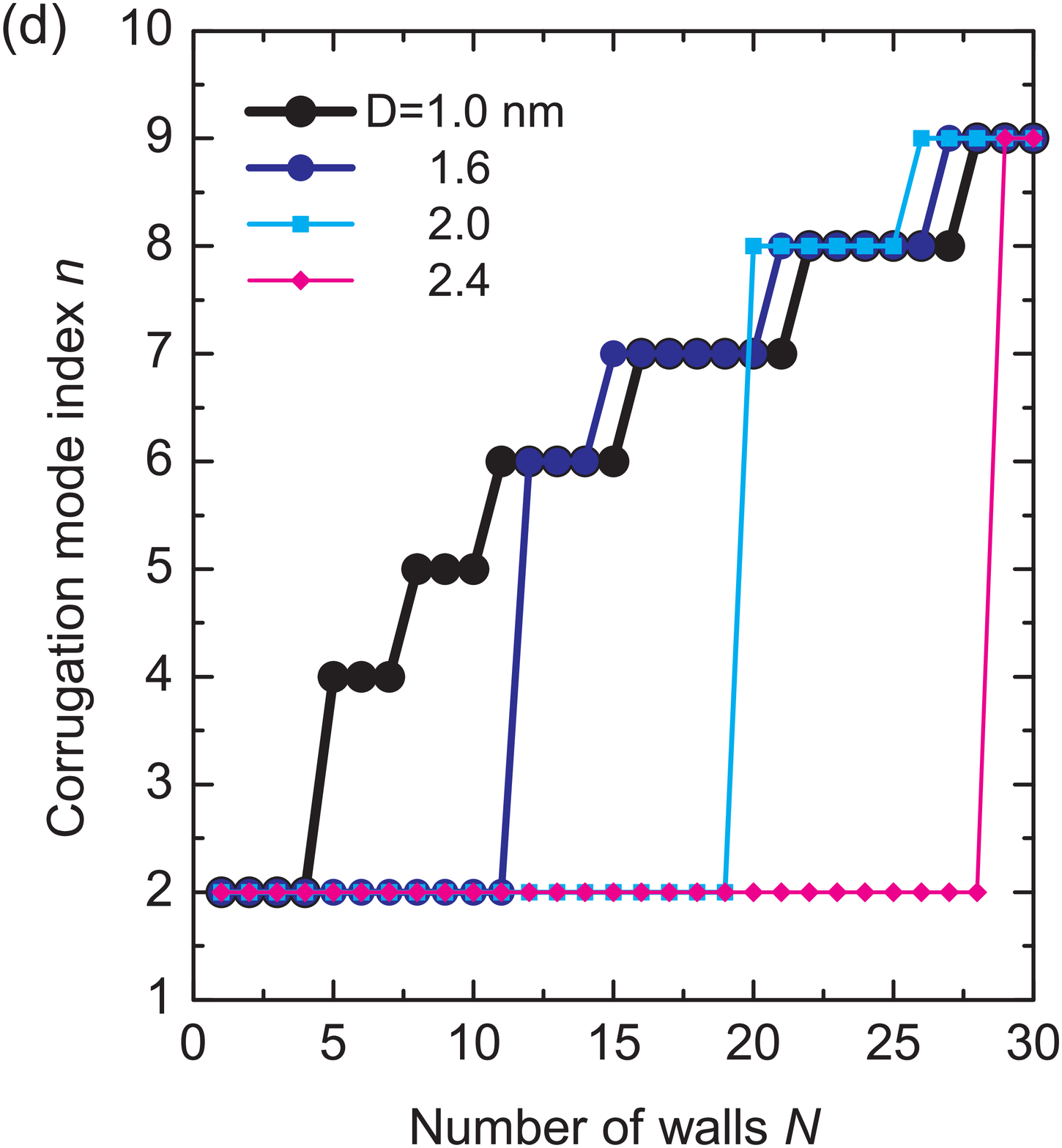}
}
\caption{
(a) Cross-sectional views of (a) elliptic $(n=2)$ and
(b) corrugated $(n=5)$ deformation modes.
The mode index $n$ indicates the wave number of the deformation mode
along the circumference.
(c) Wall-number dependence of critical pressure $p_c$. Immediately above $p_c$, the
original circular cross section of MWNTs gets radially corrugated.
(d) Stepwise increase in the corrugation mode index $n$.
Reprinted from Refs.~\cite{ShimaNTN2008,ShimaCMS2011}.}
\label{ShimaNTNFig}
\vspace*{60pt}\end{figure}

Figures \ref{ShimaNTNFig}(a) and (b) illustrate MWNT cross-sectional views
of two typical deformation modes: (a) elliptic $(n=2)$ and (b) corrugation $(n=5)$ modes.
In the elliptic mode, all constituent walls are radially deformed.
In contrast, in the corrugation mode,
outside walls exhibit significant deformation,
whereas the innermost wall maintains its circular shape.
Which mode will be obtained under pressure
depends on the number of walls, $N,$ and the core tube diameter $D$
of the MWNT considered.
In principle, larger $N$ and smaller $D$ favor
a corrugation mode with larger $n$.

Figure \ref{ShimaNTNFig}(c) shows the critical buckling pressure $p_c$
as a function of $N$ for various values of $D$.
An initial increase in $p_c$ at small $N$ (except for $D$ = 1.0 nm) is attributed to the
enhancement of radial stiffness of the entire MWNT by encapsulation.
This stiffening effect disappears with further increase in $N$,
resulting in decay or convergence of $p_c(N)$.
\color{black}
A decay in $p_c$ implies that a relatively
low pressure becomes sufficient to produce radial deformation,
thus indicating an effective ``softening" of the MWNT.
The two contrasting types of behavior, stiffening and softening, 
are different manifestations of the encapsulation effect of MWNTs.
\color{black}
It is noteworthy that practically synthesized MWNTs often show $D$ larger than those
presented in Fig.~\ref{ShimaNTNFig}(c).
Hence, $p_c(N)$ of an actual MWNT lies at several hundreds of megapascals, as estimated
from Fig.~\ref{ShimaNTNFig}(c).
Such a degree of pressure applied to
MWNTs is easily accessible in high-pressure experiments,\footnote{A radial pressure
large enough to cause corrugation
can be achieved by electron-beam irradiation \cite{KrasheninnikovJAP2010};
the self-healing nature of eroded carbon walls gives rise to a spontaneous contraction
that exerts a high pressure on the inner walls
to yield their radial corrugation \cite{ShimaPRB2010}.}
supporting the feasibility of our theoretical predictions.

Figure \ref{ShimaNTNFig}(d) shows the stepwise increases in the corrugation mode index $n$.
For all $D$, the deformation mode observed
just above $p_c$ abruptly increases from $n = 2$ to $n\ge 4$ at a certain value
of $N$, followed by the successive emergence of higher corrugation
modes with larger $n$. These successive transitions in $n$ at $N \gg 1$
originate from the mismatch in the radial stiffness of the innermost
and outermost walls. A large discrepancy in the radial stiffness of
the inner and outer walls results in a maldistribution of the deformation
amplitudes of concentric walls interacting via vdW forces,
which consequently produces an abrupt change in the observed
deformation mode at a certain value of $N$.

\begin{figure}[ttt]
\centerline{
\includegraphics[width=13.5cm]{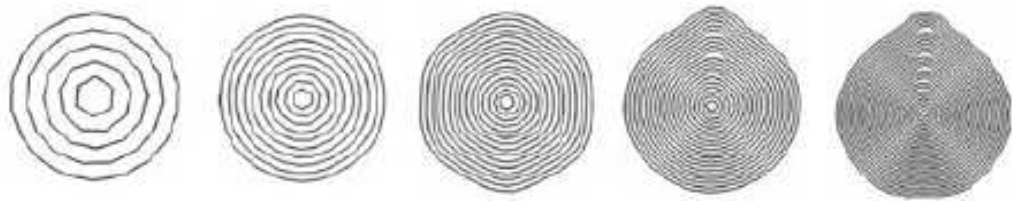}
}
\caption{
Cross-sectional views of relaxed MWNTs indexed by (2,8)/(4,16)/$\ldots$/(2$n$,8$n$).
The wall numbers $n$ are 5, 10, 15, 20, and 25 from left to right, and
all the MWNTs are 20 nm long.
Reprinted from Ref.~\cite{XHuangNanoResLett2011}.
}
\label{XHuangNanoResLett2011Fig}
\vspace*{60pt}\end{figure}

Other types of radial deformation arise when deviate the interwall spacings of MWNTs
from the vdW equilibrium distance ($\sim$ 0.34 nm) 
\cite{XHuangNanoResLett2011}.\footnote{The authors in Ref.~\cite{XHuangNanoResLett2011} say that
this conclusion was motivated by an experimental fact that
cross sections of MWNTs synthesized in the presence of nitrogen are
polygonal shapes rather than circular shapes \cite{DucatiSmall2006,KoziolAdvMater2005}.
It was argued that
the polygonization may result from the interlayer thermal contraction upon
cooling or interwall adhesion energy owing to the increased interwall commensuration \cite{DucatiSmall2006}.}
The simulations show that the cross sections stabilized at polygonal or water-drop-like
shapes, depending on the artificially expanded interwall spacings.\footnote{Prior to
structural optimization, the initial cross sections of all the MWNTs are
of circular shape and the interwall spacing is 0.359 nm,
which is 0.19 nm larger than the equilibrium spacing
of two graphene sheets (=0.34 nm).}
Figure \ref{XHuangNanoResLett2011Fig} depicts the cross-sectional configurations of
relaxed MWNTs.
It is seen that the 15-walled tube is stabilized at a polygonal cross-sectional configuration
with six rounded corners.
For the 20- and 25-walled ones, the configuration becomes asymmetric, featuring a
water-drop-like morphology.

From an engineering perspective, the tunability of the
cross-sectional geometry
may be useful in developing nanotube-based
nanofluidic \cite{MajumderNature2005,NoyNanotoday2007,WhitbyNatureNTN}
or nanoelectrochemical devices \cite{FrackowiakCarbon2001,ShimaCPL2011}
because both utilize the hollow cavity within the innermost tube.
Another interesting implication is a
pressure-driven change in the quantum transport
of $\pi$ electrons moving along the radially deformed nanotube.
It has been known that mobile electrons whose motion
is confined to a two-dimensional, curved thin layer behave
differently from those on a conventional flat plane because
of an effective electromagnetic field \cite{ShimaPRB2009,OnoPRB2009,ShimaPhysicaE2010,OnoPhysicaE2010,TairaJPCM2010A,TairaJPCM2010B}
that can affect low-energy excitations of the electrons.
\color{black}
Associated variations in the electron-phonon coupling \cite{OnoEPL2011}
and phononic transport \cite{OnoJPSJ2011}
through the deformed nano-carbon materials are also interesting and relevant to
the physics of radially corrugated MWNTs.
\color{black}

\section{Bend buckling of SWNTs}\label{sec6}

\subsection{Kink formation}\label{sec6sub1}

\begin{figure}[ttt]
\centerline{
\includegraphics[width=6.0cm]{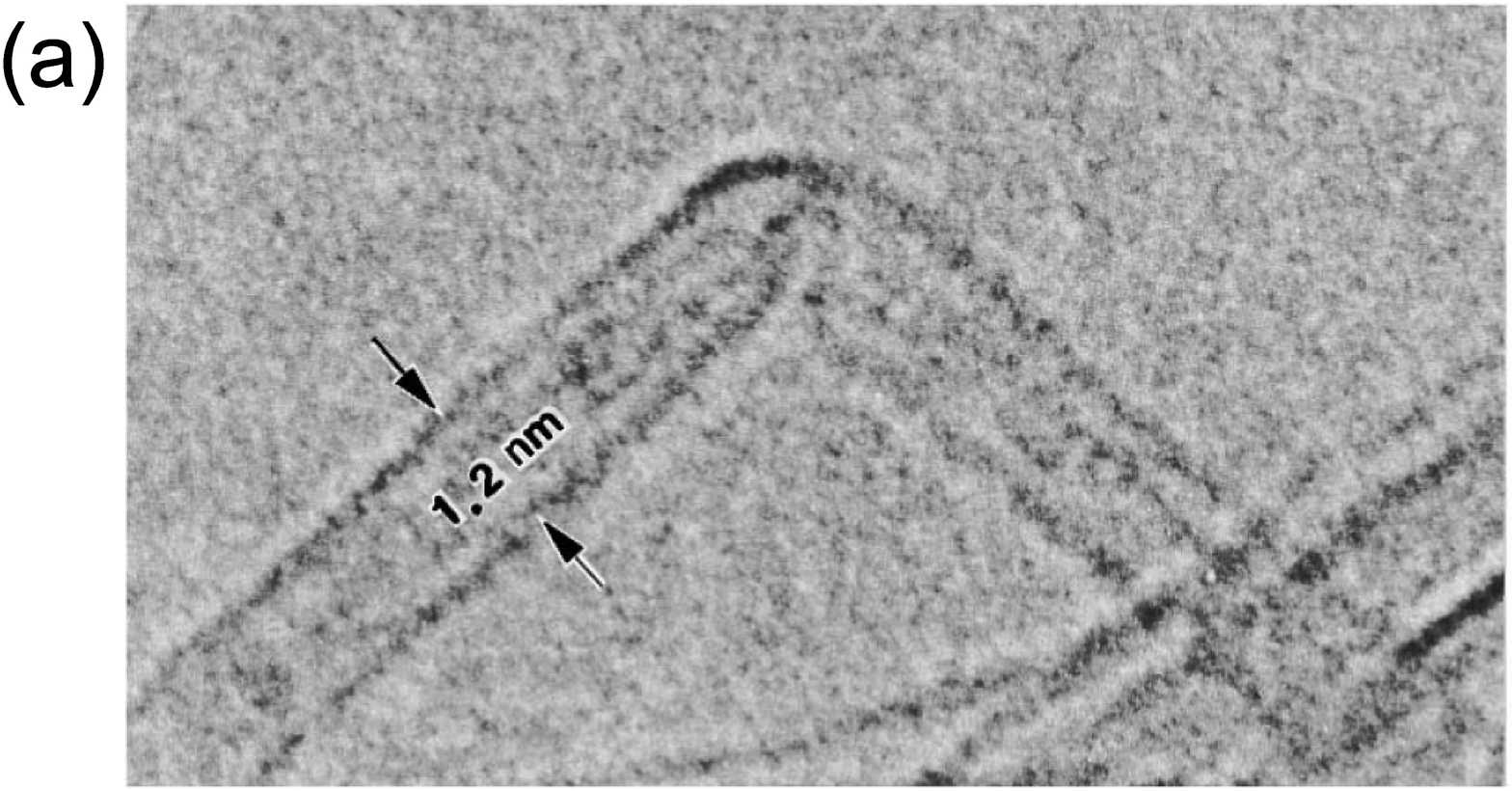}
\includegraphics[width=5.7cm]{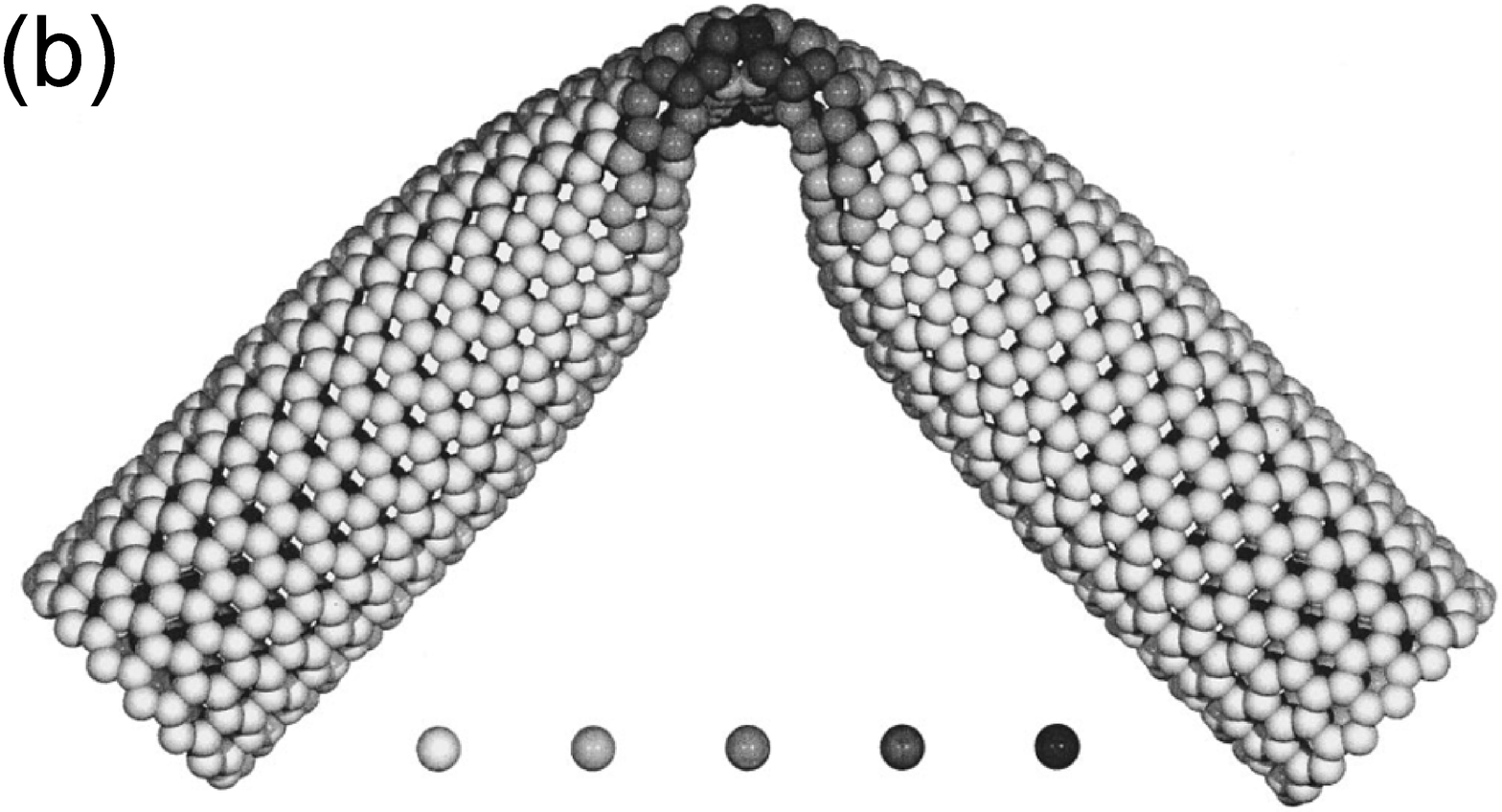}
\includegraphics[width=5.6cm]{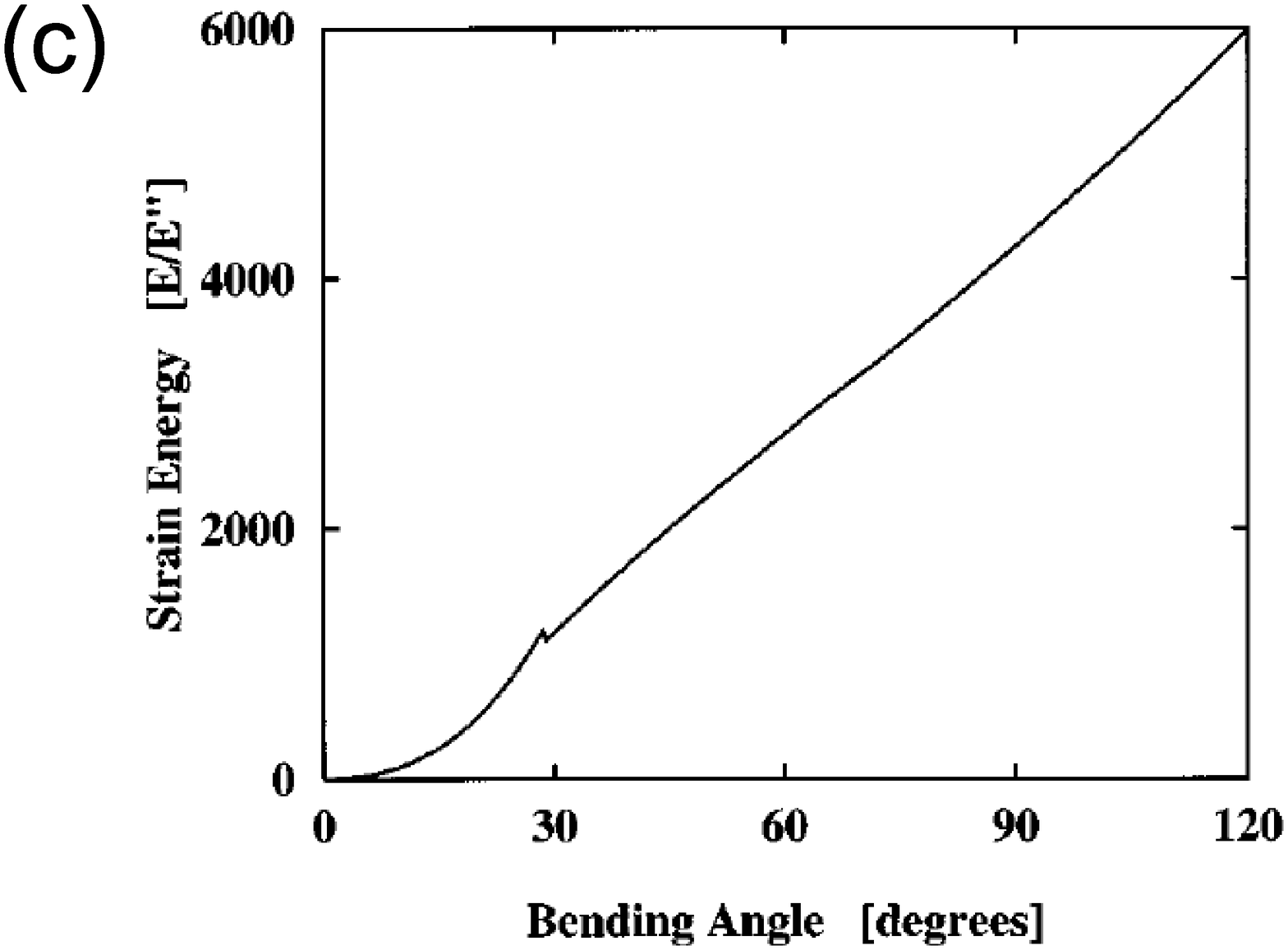}
}
\caption{(a) Kink structure formed in an SWNT with diameters of 1.2 nm under bending.
The gap between the tip of the kink and the upper wall is about 0.4 nm.
(b) Atomic structure around the kink reproduced by computer simulations.
The shaded circles beneath the tube image express
the local strain energy at the various atoms, measured relative
to a relaxed atom in an infinite graphene sheet.
The strain energy scale ranges from 0 to 1.2 eV/atom, from left to right.
(c) Total strain energy (in dimensionless units) of an SWNT
of diameter $\sim$1.2 nm as a function of the bending angle up to 120$^\circ$.
The dip at $\sim$30$^\circ$ in the curve is associated with the kink formation.
Reprinted from Ref.~\cite{IijimaJCP1996}.
}
\label{IijimaJCP1996Fig}
\vspace*{60pt}\end{figure}

The buckling of SWNTs
under bending was pioneered in 1996 \cite{IijimaJCP1996}
using high-resolution electron microscopes and molecular-dynamics (MD) simulation.
Figure \ref{IijimaJCP1996Fig}(a) shows a TEM image of a bent SWNT with a
diameter of 1.2 nm \cite{IijimaJCP1996}.
By bending an initially straight SWNT,
its outer and inner sides undergo stretching and compression, respectively.
As a result, it develops a single kink in the middle, through which
the strain energy on the compressed sides is released.
Upon removal of the bending moment, it returns to the
initial cylindrical form completely without bond breaking or defects.
This observation
clearly proves that SWNTs possess extraordinary structural elastic flexibility.

Figure \ref{IijimaJCP1996Fig}(b) presents a
computer simulated reproduction of the kink experimentally observed,
providing atomistic and energetic
information about the bending process.
The overall shape of the kink, along with
the distance of the tip of the kink from the upper wall of the tube,
is in quantitative agreement with the TEM picture of
Fig.~\ref{IijimaJCP1996Fig}(a).
The coding denotes the local strain energy
at the various atoms, measured relative
to a relaxed atom in an infinite graphene sheet.
In all simulations, the same generic
features appear: Prior to buckling ($\le 30^\circ$), the strain energy
increases quadratically with the bending angle, corresponding
to harmonic deformation [see Fig.~\ref{IijimaJCP1996Fig}(c)].
In this harmonic regime, the hexagonal
rings on the tube surface are only slightly strained
and the hexagonal carbon network is maintained.
Beyond the critical curvature, the excess strain on the compressed
side reaches a maximum and is released through the formation of a kink, which
increases the surface area of the bending side. This is accompanied
by a dip in the energy vs. bending angle curve, as
shown in Fig.~\ref{IijimaJCP1996Fig}(c).
Following the kink formation, the strain energy increases approximately
linearly until bond breaking occurs under quite large
deformation.
A similar characteristic energy-strain curve,
an initial quadratic curve followed by a linear increase,
arises in the case of axial compression \cite{YakobsonPRL1996}, as we learned in \S \ref{sec4sub2}.

\subsection{Diameter dependence}\label{sec6sub2}

\begin{figure}[ttt]
\centerline{
\includegraphics[width=7.3cm]{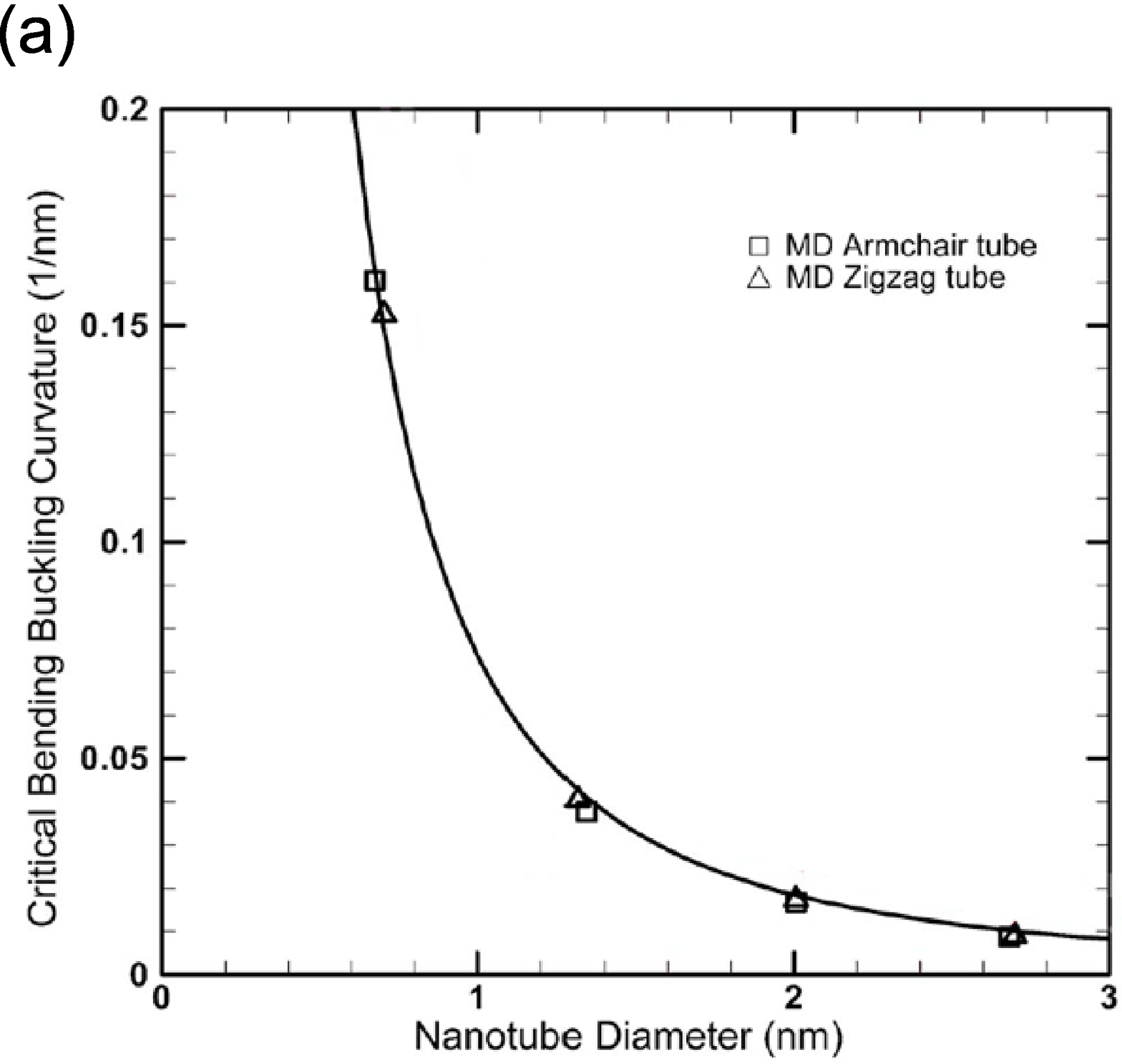}
\includegraphics[width=7.3cm]{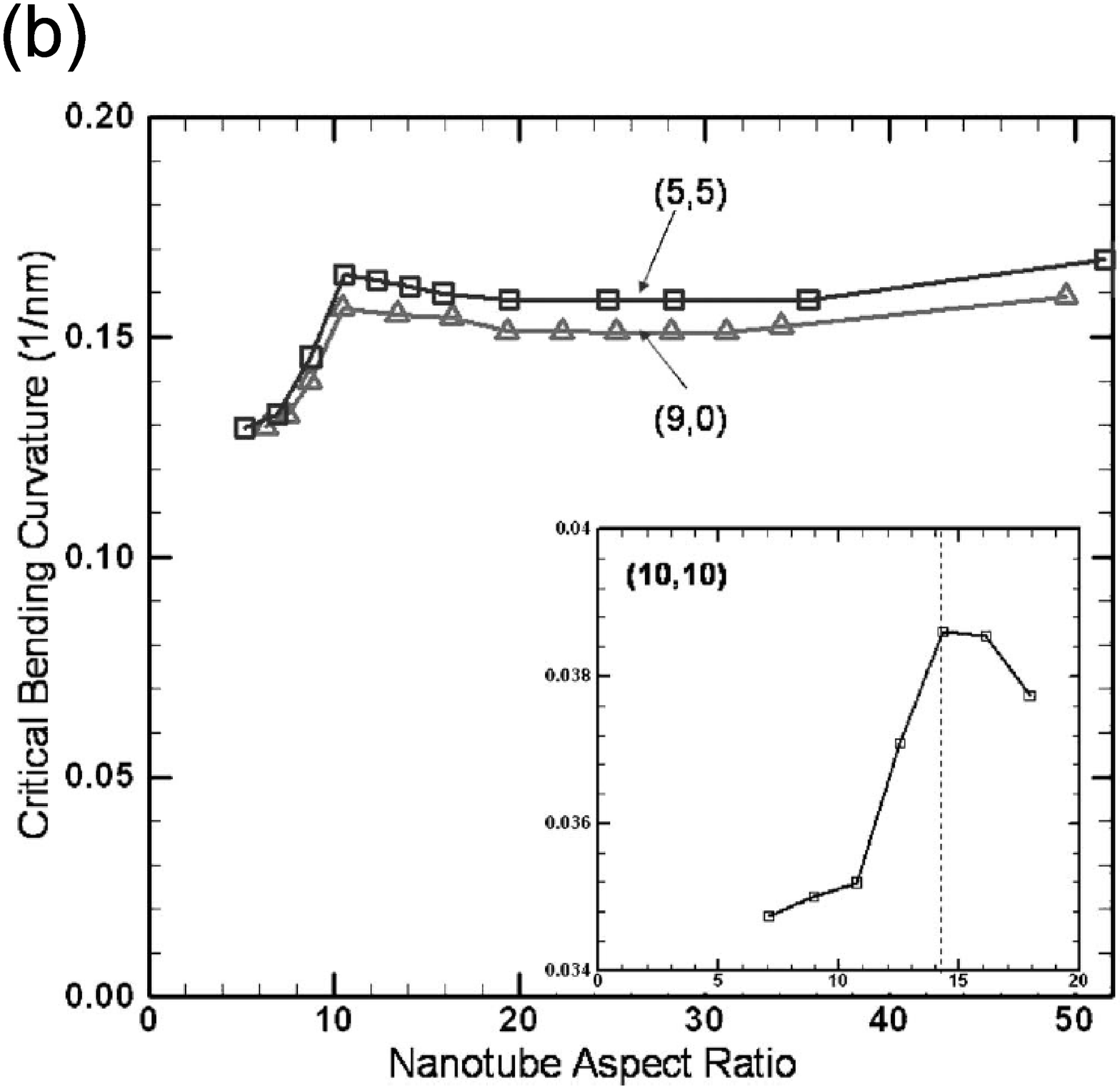}
}
\caption{
(a) Relationship between critical bending
buckling curvature $\kappa_{\rm C}$ and nanotube diameter $d$
obtained from MD analyses.
The tube length is fixed at 24 nm.
(b) The length/diameter $(L/d)$ aspect ratio
dependence of $\kappa_{\rm C}$ for the tube chiralities of (5,5), (9,0), and (10,10).
Reprinted from Ref.~\cite{GCaoPRB2006}.
}
\label{GCaoPRB2006Fig}
\vspace*{60pt}\end{figure}

The geometrical size is a crucial factor for determining buckling
behaviors of SWNTs under bending.
For instance, those with a small diameter can sustain a large bending angle
prior to buckling, and vice versa \cite{GCaoPRB2006}.
Figure \ref{GCaoPRB2006Fig}(a) shows MD simulation data,
which show a monotonic increase in the critical curvature $\kappa_{\rm C}$ for 
with a reduction in the nanotube diameter $d$.
The relationship between $\kappa_{\rm C}$ and $d$ can be fitted as \cite{YakobsonPRL1996,GCaoPRB2006}
$\kappa_{\rm C} \propto d^{-2}$,
which holds regardless of the nanotube chirality.

In addition to the diameter dependence,
the critical curvature of SWNTs is affected by the length/diameter ($L$/$d$) aspect ratio.
It follows from Fig.~\ref{GCaoPRB2006Fig}(b) that \cite{GCaoPRB2006}
$\kappa_{\rm C}$ is almost constant for sufficiently long nanotubes such that $10< L/d < 50$ or more,
whereas it drops off for short nanotubes satisfying $L/d < 10$.
The critical aspect ratio that separate the two regions
is sensitive to the tube diameter [as implied by the inset of Fig.~\ref{GCaoPRB2006Fig}(b)],
but it is almost independent of
chirality.

\subsection{Transient bending}\label{sec6sub3}

\begin{figure}[ttt]
\centerline{\includegraphics[width=9.0cm]{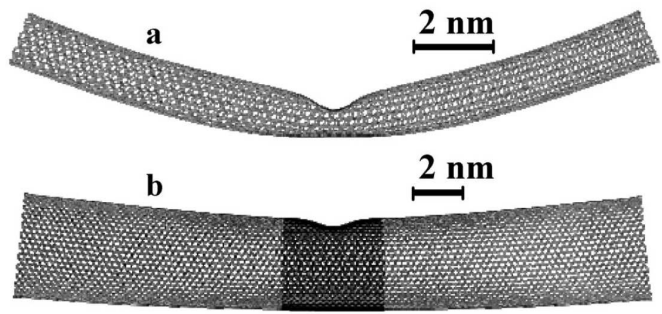}
}
\caption{Predicted shape of SWNTs just after buckling,
based on MD simulations for
(a) a 15.7-nm-long (10, 10) SWNT at the bending angle $\theta = 43^\circ$
and (b) a 23.6-nm-long (30, 30) SWNT at $\theta = 23^\circ$.
Note the difference in scale.
Reprinted from Ref.~\cite{KutanaPRL2006}.
}
\label{KutanaPRL2006Fig1}
\vspace*{60pt}\end{figure}

\begin{figure}[hhh]
\centerline{
\includegraphics[width=7.5cm]{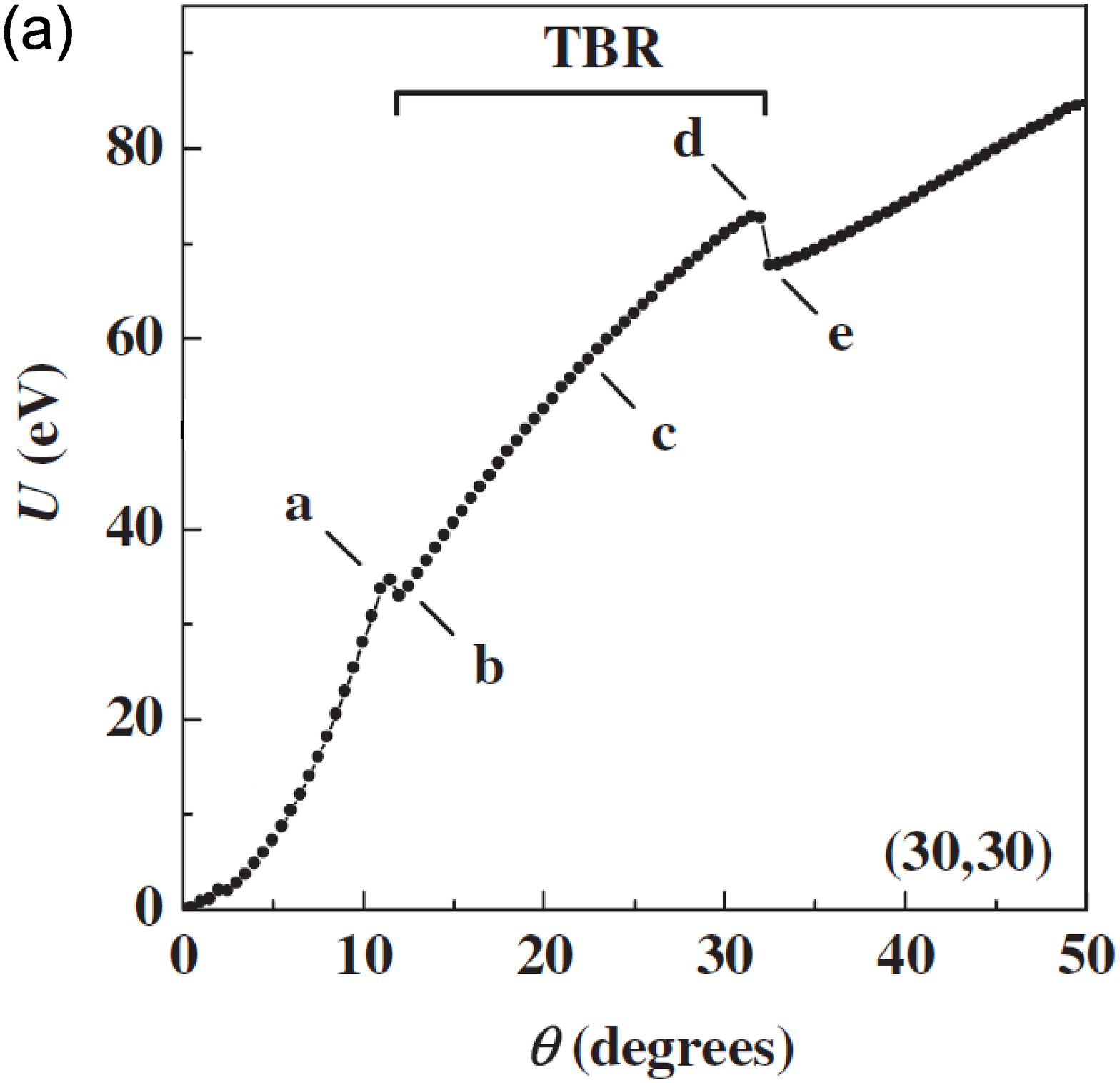}
\hspace*{1cm}
\includegraphics[width=5.0cm]{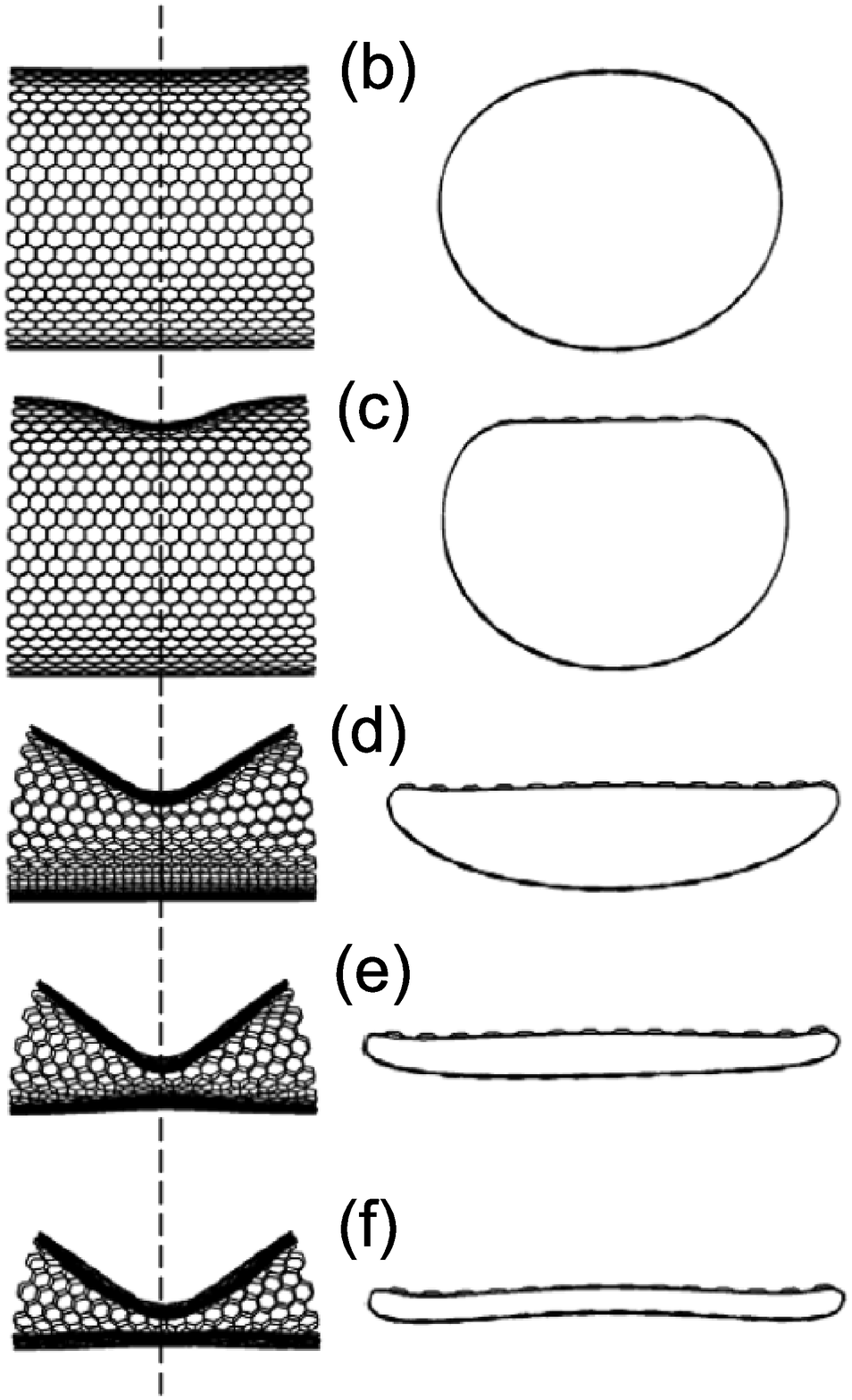}
}
\caption{
[Left] Deformation energy $U$ for a 23.6-nm-long (30,30) SWNT as a function of the bending
angle $\theta$. The symbols a--e attached to the curve
indicate the points for which the tube
shape and cross section at the buckling point are shown in the images (b)--(f) on the right. 
``TBR" denotes the transient bending regime.
Reprinted from Ref.~\cite{KutanaPRL2006}.
}
\label{KutanaPRL2006Fig2and3}
\vspace*{60pt}\end{figure}

We have learned in \S \ref{sec4} that
for relatively thin SWNTs,
the buckling is characterized by a discontinuity
in the energy curve [Fig.~\ref{IijimaJCP1996Fig}(c)].
Across the buckling point, the bending angle dependence of the deformation energy
changes suddenly from being quadratic  in the prebuckling regime
to  linear  in the postbuckling regime.
This is, however, not the case for larger diameter nanotubes.
As the diameter is increased,
a second discontinuity appears in the strain-energy curve
at a larger bending angle than the first one \cite{KutanaPRL2006}.
The origin of the two discontinuities can be accounted for by inspection of
Fig.~\ref{KutanaPRL2006Fig1}.
The bottom image [Fig.~\ref{KutanaPRL2006Fig1}(b)] shows
a just-buckled wall of a (30, 30) SWNT corresponding to the first discontinuity
in the energy curve, in which the buckled side is far from the opposite side.
Therefore, more bending is required
to bring the two sides close enough, as observed in Fig.~\ref{KutanaPRL2006Fig1}(a),
which results in the second discontinuity.

The thick-nanotube's buckling behavior mentioned above
is illustrated in Fig.~\ref{KutanaPRL2006Fig2and3}(a),
where the deformation energy $U$ is plotted as a function of
bending angle $\theta$ for a (30, 30) SWNT \cite{KutanaPRL2006}.
Three distinct deformation
regimes are observed, clearly
separated by two discontinuities at $\theta = 12^\circ$ and $32^\circ$.
In the initial elastic regime, $U$ exhibits a quadratic
dependence on $\theta$, whereas the cross section
experiences progressive ovalization as the bending angle
increases, culminating to the shape in Fig.~\ref{KutanaPRL2006Fig2and3}(b).
The buckling event is marked by an abrupt transition from the oval
cross section to one with the flat top shown in Fig.~\ref{KutanaPRL2006Fig2and3}(c).
As the bending angle increases beyond the first discontinuity
[{\it i.e.,} during the transient bending regime (TBR)
indicated in
Fig.~\ref{KutanaPRL2006Fig2and3}(a)],
the flat portion of the top wall expands continuously across the
nanotube [Figs.~\ref{KutanaPRL2006Fig2and3}(d) and (e)].\footnote{Interestingly,
a buckled SWNT in the TBR is fully reversible. If bending
is stopped before the second discontinuity occurs, unbending
recovers the cross-sectional shapes at the buckling point.}
As a result,
the top-to-bottom wall distance decreases gradually, reducing
the tube cross section at the buckling site. 
The deformation energy curve in the TBR is no longer
quadratic; in fact, the exponent becomes less than unity.
When the approaching opposite walls reach the
vdW equilibrium distance
of 0.34 nm [Fig.~\ref{KutanaPRL2006Fig2and3}(f)],
the cross section collapses, forming the kink, and a second
discontinuity is observed.


\section{Bend buckling of MWNTs}\label{sec7}

\subsection{Emergence of ripples}\label{sec7sub1}

\begin{figure}[ttt]
\centerline{
\includegraphics[width=6.0cm]{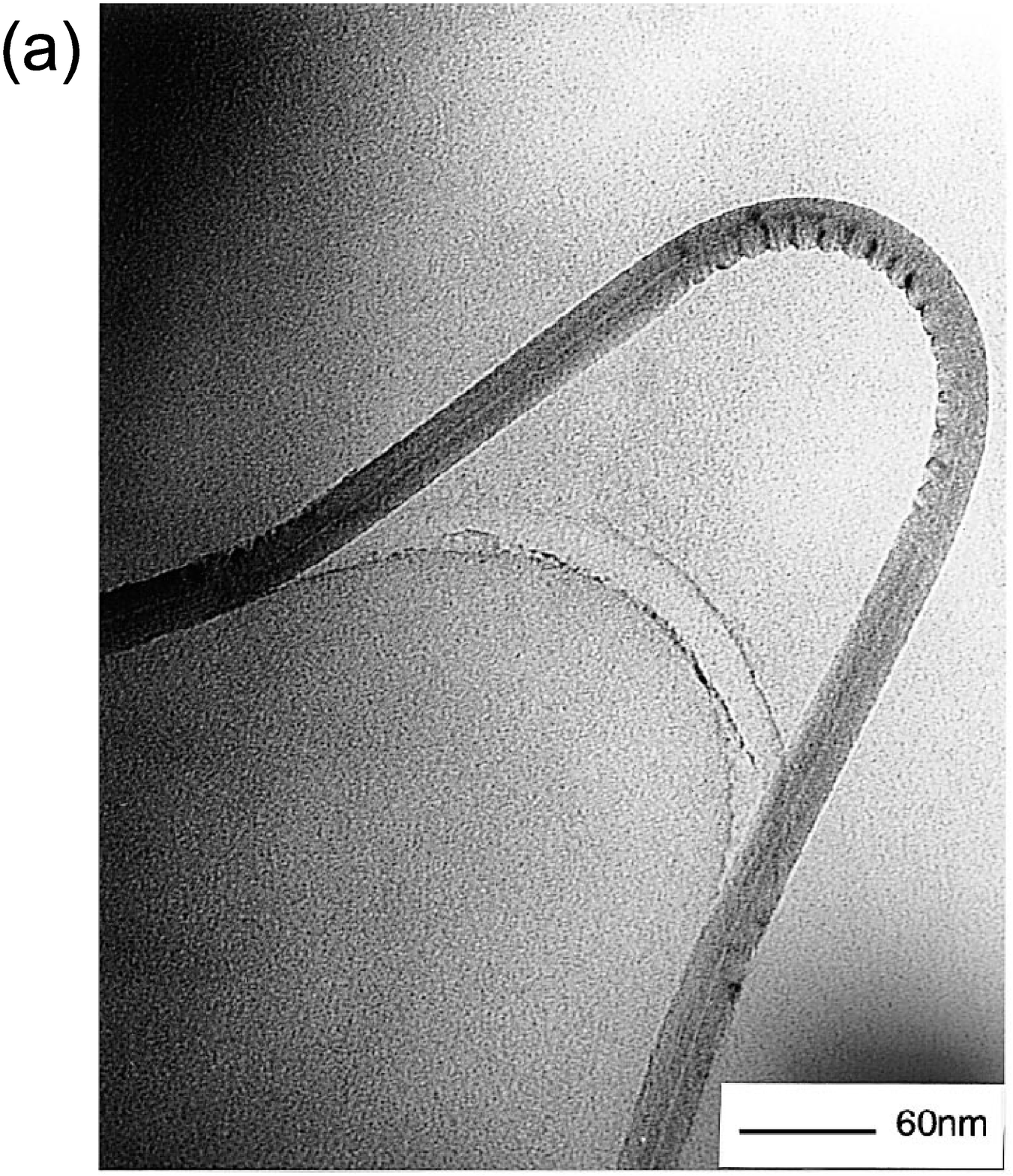}
\includegraphics[width=7.9cm]{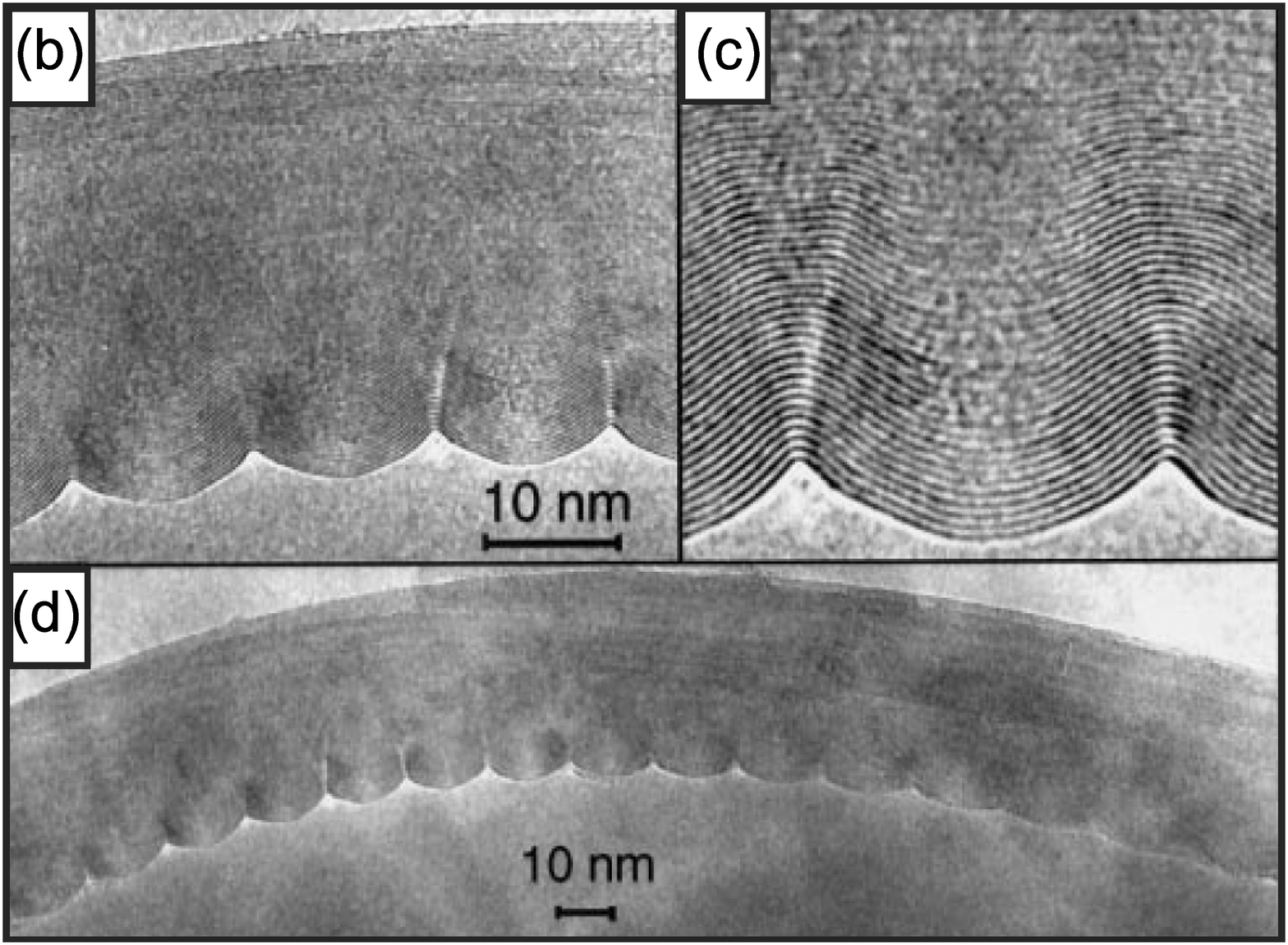}
}
\caption{
(a) Under high bending, MWNTs form kinks
on the internal (compression) side of the bend.
(b)--(d) High-resolution  TEM image of a bent nanotube
(with a radius of curvature of $\sim$400 nm), showing the characteristic wavelike
distortion. The amplitude
of the ripples increases continuously from the center of the tube to the
outer layers of the inner arc of the bend.
Reprinted from Refs.~\cite{LouriePRL1998,PoncharalScience1999}.
}
\label{LouriePRL1998Fig2undPoncharalScience1999Fig3bcd}
\vspace*{60pt}\end{figure}

Following the discussions on SWNTs in \S \ref{sec6},
we look into the bend buckling of MWNTs.
The difference in the mechanical responses between SWNTs
and MWNTs lies in the presence of vdW interactions between
the constituent carbon layers.
Apparently, thicker MWNTs with tens of concentric walls
seem stiffer than few-walled, thin MWNTs against bending,
since the inner tubes of MWNTs may reinforce the outer tubes via the vdW interaction.
However, the contrary occurs.
In fact, whereas MWNTs with small diameter exhibited
a bending stiffness of around 1 TPa, those with larger diameter were much
more compliant, with a stiffness of around 0.1 TPa \cite{PoncharalScience1999}.
This dramatic reduction in the bending stiffness was attributed to the
so-called
rippling effect, {\it i.e.,}
the emergence of a wavelike distortion on
the inner arc of the bent nanotube
\cite{RuoffCarbon1995,KuzumakiPhilMagA1998,LouriePRL1998,BowerAPL1999,ArroyoPRL2003,
PantanoPRL2003,TChangJAP2006,XYLiPRL2007,AriasPRL2008,
ArroyoJMPS2008,XHuangAPL2008,JZouJAP2009,NikiforovAPL2010}.

Figure \ref{LouriePRL1998Fig2undPoncharalScience1999Fig3bcd}(a)
presents a clear example of the rippled MWNT structure \cite{LouriePRL1998}.
The tube diameter is $\sim$31 nm and it is subjected to a radius of bending curvature
of $\sim$400 nm.
Enhanced images at the ripple region \cite{PoncharalScience1999} are also displayed in
Figs.~\ref{LouriePRL1998Fig2undPoncharalScience1999Fig3bcd}(b)--(d),
though they are not identical to the specimen in Fig.~\ref{LouriePRL1998Fig2undPoncharalScience1999Fig3bcd}(a).
The amplitude of the ripple increased gradually from
inner to outer walls,
being essentially zero for the innermost core tube
to about 2 to 3 nm for the outermost wall.
Such rippling deformation induces
a significant reduction in the bending modulus,
as has been explained theoretically
by solving nonlinear differential equations \cite{JZLiuPRL2001}.

\subsection{Yoshimura pattern}\label{sec7sub2}

\begin{figure}[ttt]
\centerline{\includegraphics[width=8.5cm]{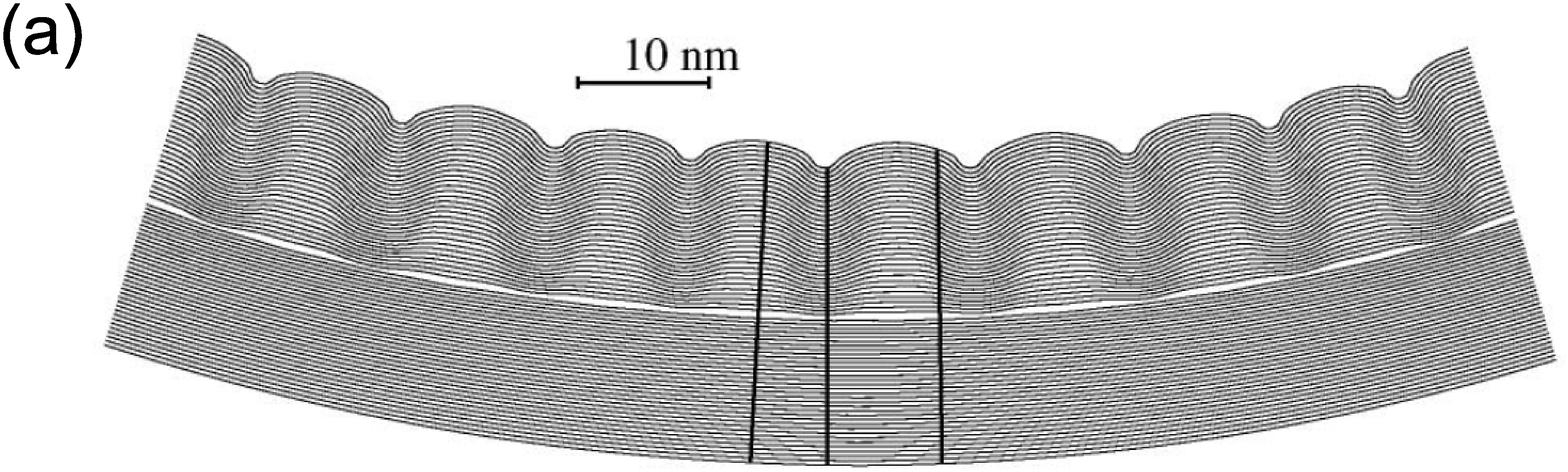}}
\centerline{\includegraphics[width=8.5cm]{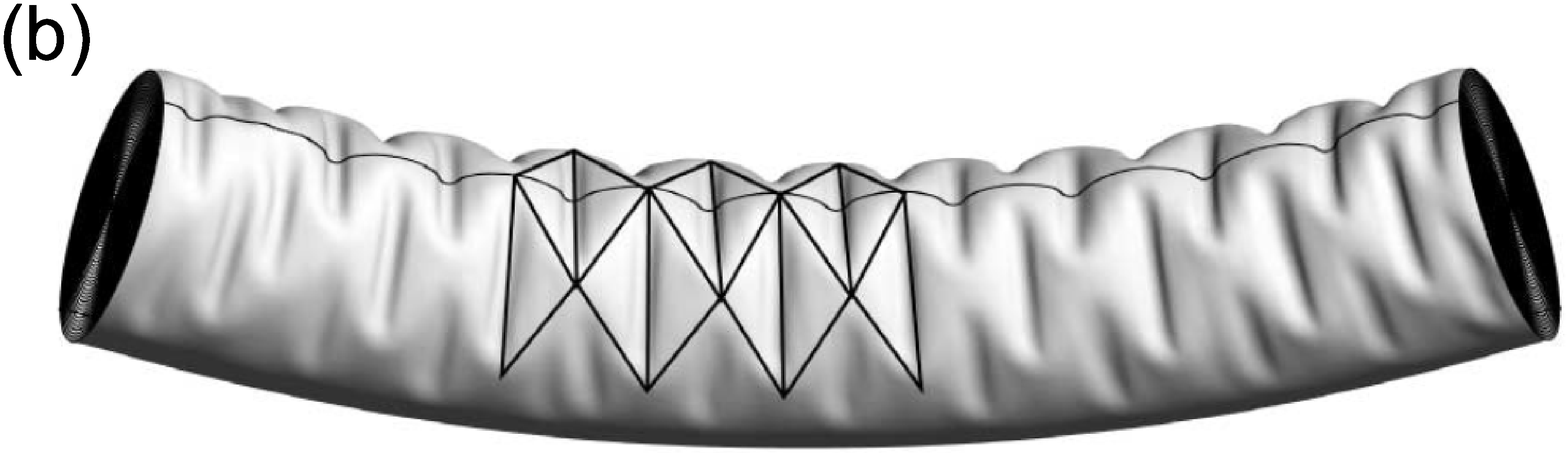}}
\caption{Rippling of a 34-walled carbon nanotube: (a) longitudinal
section of the central part of the simulated nanotube
and (b) the morphology of the rippled MWNT reminiscent of the Yoshimura pattern.
Highlighted are the ridges and furrows, as well as the
trace of the longitudinal section.
Reprinted from Ref.~\cite{ArroyoPRL2003}.
}
\label{ArroyoPRL2003Fig1}
\vspace*{60pt}\end{figure}

Precise information about the membrane profile and the energetics of
the rippling deformation, which are unavailable in experiments,
can be extracted from large-scale computer simulations.
Figure \ref{ArroyoPRL2003Fig1}(a) shows a longitudinal cross section of the
equilibrium configuration \cite{ArroyoPRL2003}. 
This image is the computational analog of the TEM slices of rippled, thick nanotubes
reported in the literature \cite{KuzumakiPhilMagA1998,PoncharalScience1999}.
The simulations
reproduce very well the general features of the observed
rippled nanotubes: nearly periodic wavelike distortions,
whose amplitudes vanish for the inner tubes and
smoothly increase toward the outer layer.

A remarkable finding in the simulations is that the rippling deformation
closely resembles the Yoshimura pattern\footnote{The Yoshimura pattern
is a special kind of surface deformation mode occurring in
thin-walled cylindrical shells subjected to large lateral load.
It is named after Prof.~Yoshimura \cite{YoshimuraPatterns},
a Japanese theoretician of  fracture mechanics,
and its profile is characterized by a periodic diamond-like corrugation.
}
(a diamond buckling pattern).
We can see that the rippling profile in Fig.~\ref{ArroyoPRL2003Fig1}(b)
consists not of a simple linear sequence of kinks
but of a diamond-like configuration of kinks on the compressed side.
Such a Yoshimura pattern
is well known as
 characterizing the postbuckling behavior of cylindrical elastic shells
on a conventional macroscopic scale.
The pattern has the interesting geometric property
of being a nearly isometric mapping of the undeformed
surface, at the expense of creating sharp ridges and furrows.

The rippling deformation,
peculiar to thick MWNTs,
is a consequence of the interplay
between the strain-energy relaxation
and the vdW energy increment.
As intuitively understood, the 
\color{black}
low
\color{black}
bending 
\color{black}
rigidity
\color{black}
of individual 
\color{black}
graphitic sheets,
\color{black}
relative to 
\color{black}
their large
\color{black}
in-plane stiffness,
\color{black}
makes it possible to release effectively a significant amount of
\color{black}
the membrane strain energy at the
expense of slight flexural energy.
As a result, rippled MWNTs have a significantly lower
strain energy than uniformly bent MWNTs.

\begin{figure}[ttt]
\centerline{\includegraphics[width=7.5cm]{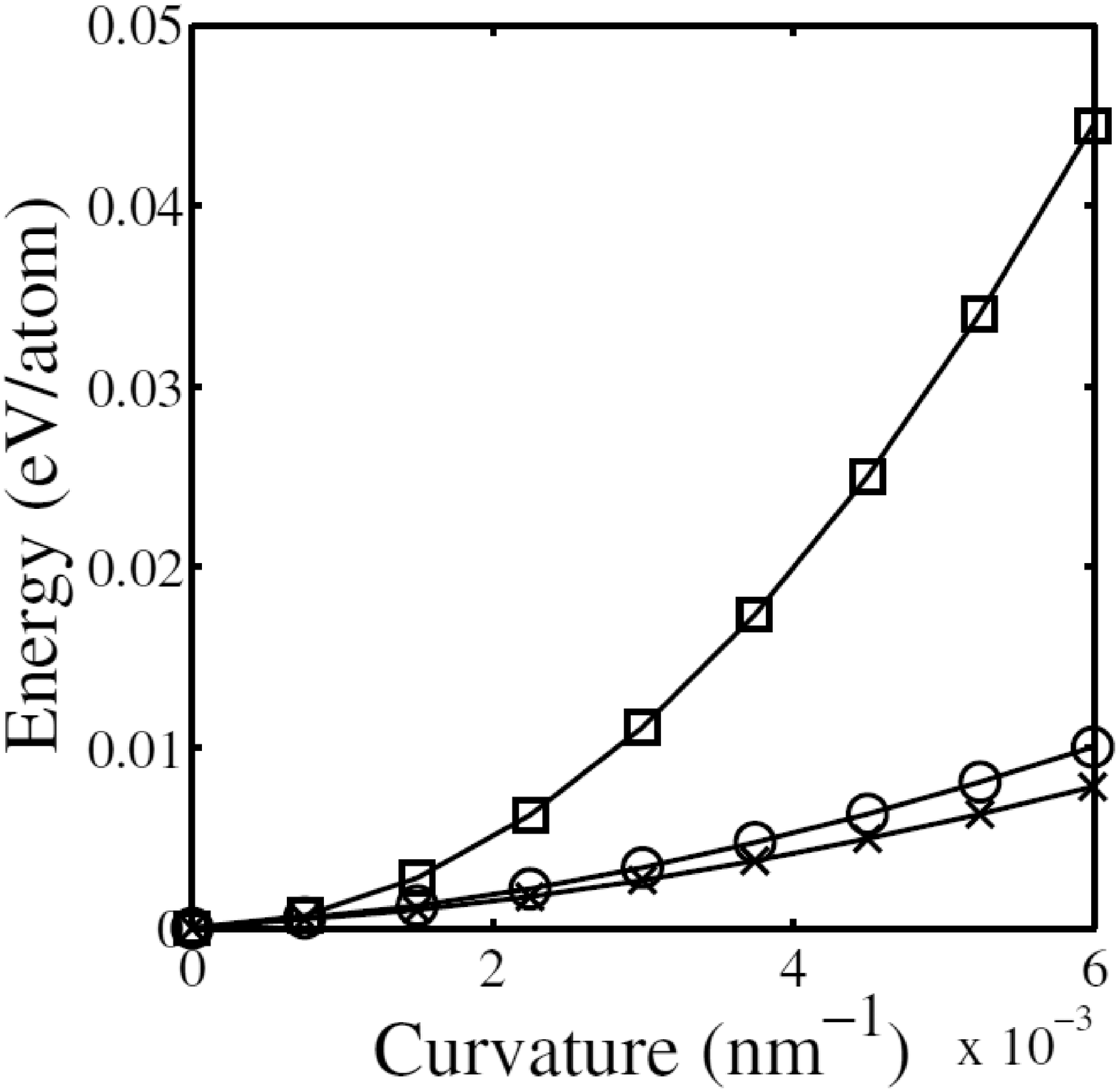}}
\caption{Energy curves for a bent 34-walled nanotube with respect to the bending curvature.
Shown are the strain energy for fictitiously uniform bending (squares),
the strain energy for actually rippled deformation (crosses),
and the total energy ({\it i.e.}, sum of the strain energy and the vdW one)
for rippled deformation (circles).
Reprinted from Ref.~\cite{ArroyoPRL2003}.
}
\label{ArroyoPRL2003Fig2}
\vspace*{60pt}\end{figure}

Figure \ref{ArroyoPRL2003Fig2} shows the energy of bent MWNTs as a function
of the bending curvature for a 34-walled nanotube \cite{ArroyoPRL2003}.
When the nanotube is uniformly bent, the strain energy grows quadratically
with respect to the curvature.
For such a uniform bending, the vdW energy
gives almost no contribution
to deformation, and therefore, the total energy
also follows a quadratic law. However, the actual behavior
of the system greatly deviates from this ideal
linearly elastic response.
As can be observed in Fig.~\ref{ArroyoPRL2003Fig2},
the rippling deformation
leads to much lower values of strain energy
and an increase in vdW energy.
The evolution of the total energy $E_{\rm tot}$
with respect to curvature radius $R$ is very accurately fitted by
$E_{\rm tot} \propto R^{-a}$ with $a=1.66$;
this response differs from that predicted by atomistic
simulations of SWNTs or small, hollow MWNTs, both of which exhibit
an initial quadratic growth ($a=2$) in the elastic regime,
followed by a linear growth ($a=1$) in the postbuckling regime.
The results in Figs.~\ref{ArroyoPRL2003Fig1} and \ref{ArroyoPRL2003Fig2}
evidence the failure of the linear elasticity and linearized
stability analysis to explain the observed well-defined postbuckling
behavior ($1<a<2$) for thick nanotubes,
implying the need for a new theoretical framework
based on  nonlinear mechanics.\footnote{Interestingly,
the energetics of MWNT bucking under bending
are altered by inserting cross-linking ({\it i.e.,} sp$^3$ covalent bonding)
between adjacent walls, as presented in Refs.~\cite{XHuangAPL2010,DuchampJAP2010}.}


\section{Twist buckling}\label{sec8}

\subsection{Asymmetric response of SWNTs}\label{sec8sub1}

Similarly to bending situations,
SWNTs under torsion exhibit a sudden morphological change
at a critical torque,
transforming into a straight-axis helical shape.
The crucial difference from bending cases is that, under torsion,
the critical buckling torque of SWNTs  depends on the loading
direction, {\it i.e.,} whether the tube is twisted in a right-handed or left-handed manner
 \cite{TChangAPL2007,JGengPRB2006}.
This load-direction dependence originates
from the tube chirality, which breaks the rotational symmetry about the tube axis.
For example, the twisting failure strain of chiral SWNTs
in one rotational direction may even be 25\% lower than
that in the opposite direction \cite{JGengPRB2006}.
Moreover,  symmetry breaking causes coupling between axial tension and torsion,
giving rise to an axial-strain-induced torsion of chiral SWNTs \cite{HLiangPRL2006}.
This intriguing coupling effect shows
that a chiral SWNT can convert motion between rotation
and translation, thus promising a potential utility
of chiral SWNTs as electromechanical device components.

\begin{figure}[ttt]
\centerline{\includegraphics[width=11.0cm]{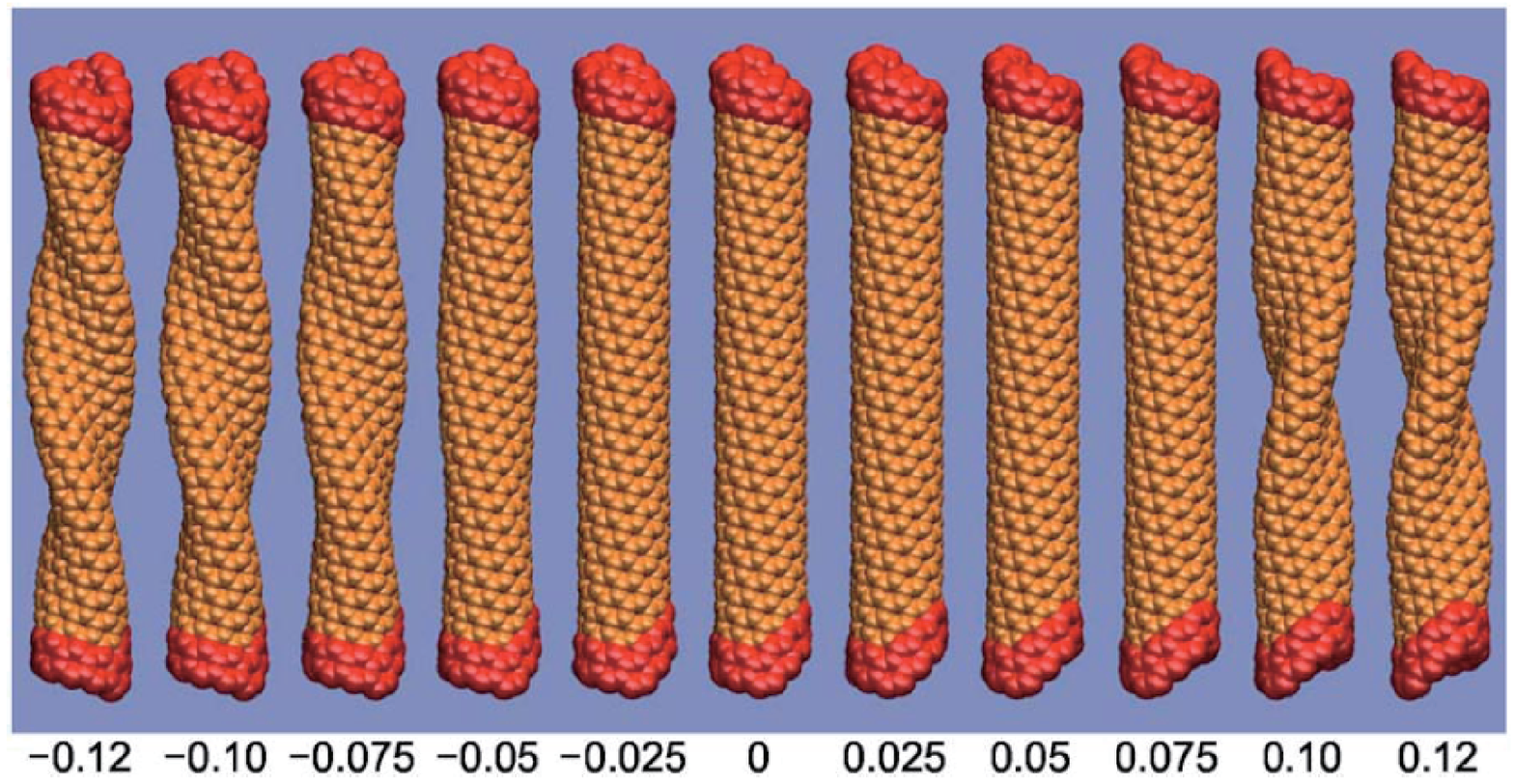}}
\caption{Morphological changes for a $(8,3)$ nanotube under torsion.
\color{black}
Applied strain and its direction are indicated beneath the diagram;
the digit 0.05, for example, corresponds to the strain of 5\%,
and the sign $+$ ($-$) indicates right(left)-handed rotation.
\color{black}
Under right-handed rotation,
the tube buckles at
a critical buckling strain $\gamma_{\rm cr}$
=7.6\%, whereas it buckles at $\gamma_{\rm cl}=4.3\%$
under left-handed rotation.
Reprinted from Ref.~\cite{TChangAPL2007}.
}
\label{TChangAPL2007Fig1}
\vspace*{60pt}\end{figure}

\begin{figure}[ttt]
\centerline{\includegraphics[width=8.5cm]{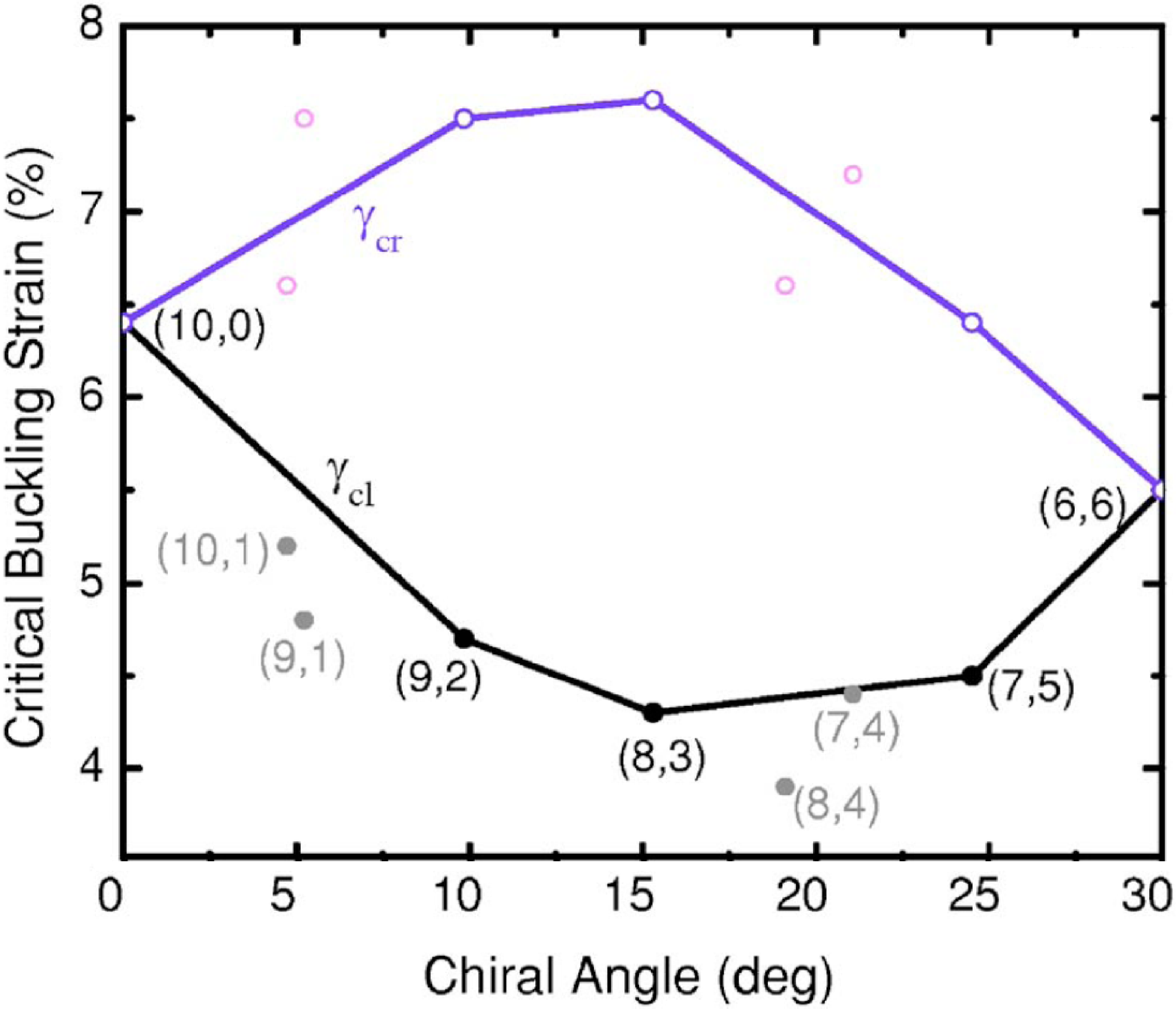}}
\caption{Critical buckling shear strains as a function of tube chirality.
Some additional data for SWNTs with slightly larger or
smaller diameters are also presented for reference.
Reprinted from Ref.~\cite{TChangAPL2007}.
}
\label{TChangAPL2007Fig3}
\vspace*{60pt}\end{figure}

The effect of structural
details on buckling of a torsional
SWNT was explored using
MD calculations \cite{TChangAPL2007}.
Figure \ref{TChangAPL2007Fig1} shows morphology changes of an (8,3) SWNT under torsion.
Its torsional deformation depends significantly on the loading direction.
Under  right-handed rotation,
the tube buckles at
a critical buckling strain of $\sim$7.6\%, which is significantly larger than that ($\sim$4.3\%)
under left-handed rotation.
Figure \ref{TChangAPL2007Fig3} summarizes a systematic computation \cite{TChangAPL2007} of
the critical buckling strains in
both twisting ($\gamma_{\rm cr}$) and untwisting ($\gamma_{\rm cl}$) directions as a function of tube chiral angle;
a loading-direction-dependent torsional response of chiral tubes is clearly observed. 
Special attention should
be paid to the fact that the maximum difference 
between $\gamma_{\rm cr}$ and $\gamma_{\rm cl}$ is  up to 85\%.
This clear difference in the mechanical response
suggests particular caution in the use of
carbon nanotubes as torsional components (e.g., oscillators and springs)
of nanomechanical devices \cite{KarniNatureNTN2006,HallNatureNTN2007,NagapriyaPRL2008,HallNanoLett2008}.

\begin{figure}[ttt]
\centerline{
\includegraphics[width=8cm]{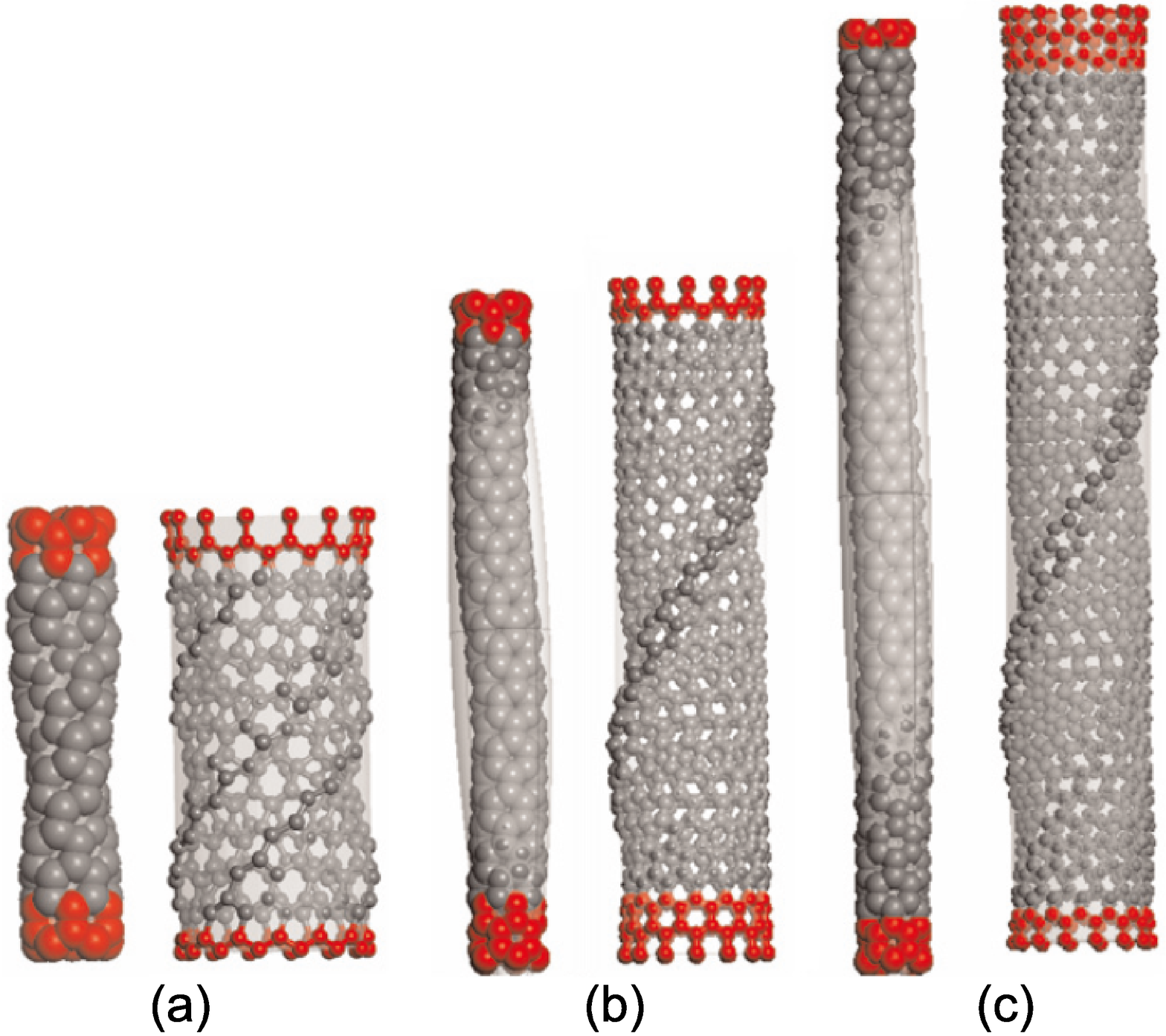}
\hspace*{2cm}
\includegraphics[width=4.3cm]{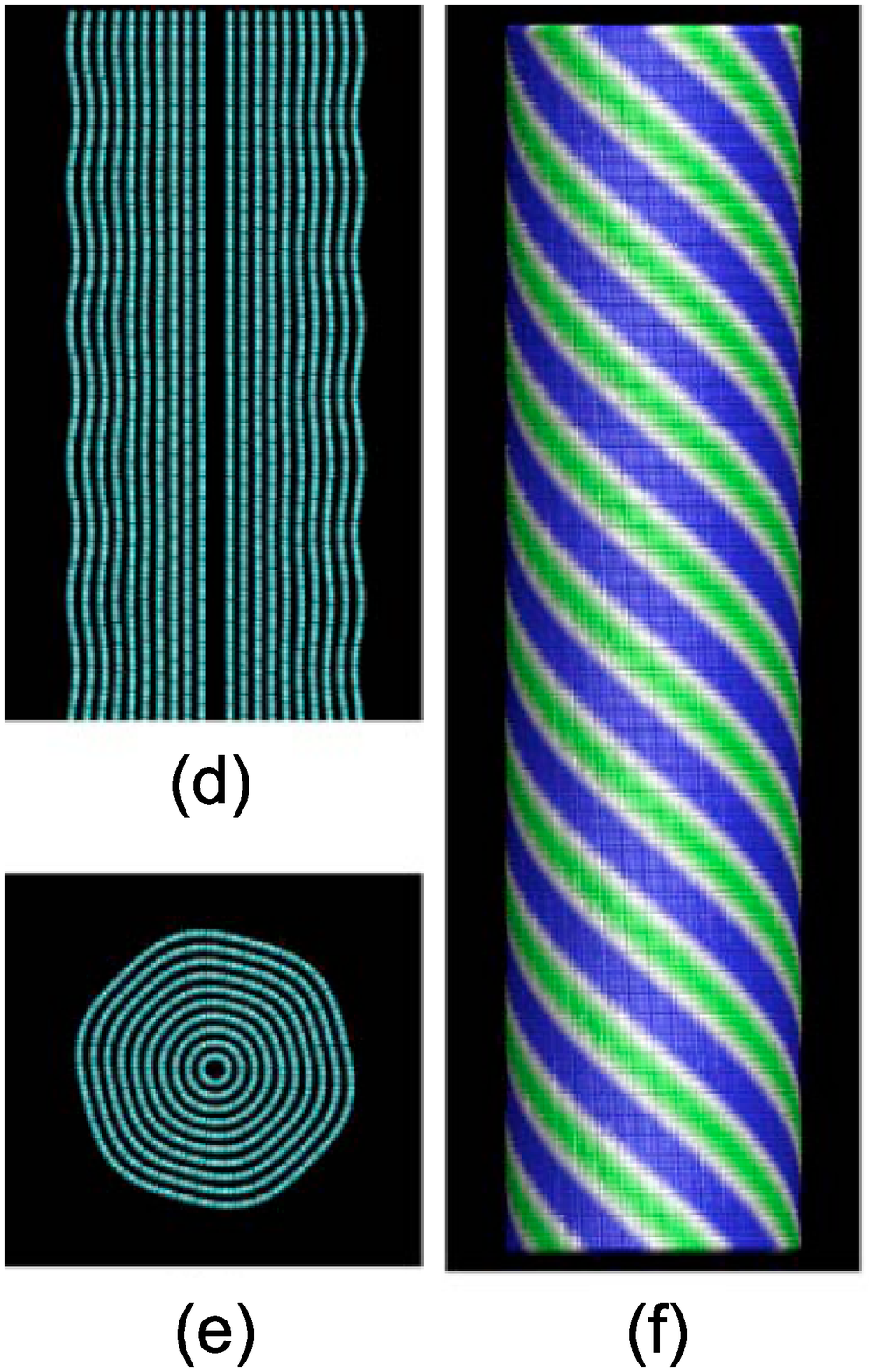}
}
\caption{
(a)--(c) Helical buckling of (5,0)@(14,0) DWNTs
with  lengths (a) $L =$ 1.095 nm, (b) 4.45 nm, and (c) 6.97 nm.
For each tube length, the inner wall shows ordinary buckled patterns,
whereas the outer wall exhibits nontrivial buckling modes associated with local rims.
(d)--(f) Helical rippling deformation of a 10-walled nanotube
(5,5)@$\dots$@(50,50) of 34 nm in length and 3.4 nm in radius.
(d) Longitudinal view. (e) Cross-sectional view.
(f) Deformation map, with green for ridges and blue for furrows.
Reprinted from Refs.~\cite{QWangCarbon2008,JZouJAP2009}.
}
\label{QWangCarbon2008Fig2undJZouJAP2009Fig4}
\vspace*{60pt}\end{figure}

\subsection{Nontrivial response of MWNTs}\label{sec8sub2}

In contrast to SWNTs cases, fewer studies on MWNTs
under torsion were reported because of their complex structures
and computational costs.
For twisted DWNTs,
MD simulations have revealed \cite{QWangCarbon2008,BWJeongCarbon2010}
a nontrivial buckling mode involving a few thin, local rims
on the outer tube while the inner tube shows a helically aligned buckling mode
[Figs.~\ref{QWangCarbon2008Fig2undJZouJAP2009Fig4}(a)--(c)].
These distinct buckling modes of the two concentric tubes imply that
a conventional continuum approximation in which it is postulated
that the buckling modes of all the constituent tubes have the same shape
fails for analyzing the torsional responses of DWNTs.

When increasing the number of constituent walls to far more than two,
we acquire
torsional rippling deformations \cite{DBZhangACSNano2010}.\footnote{During  torsional rippling,
the innermost core tube stores a very high strain energy
despite the nearly zero rippling amplitude.
Owing to the strong confinement,
the strain energy in this layer cannot be released via
rippling, which may lead to bond breaking and subsequent
brittle cracking when the torsional deformation continues to
increase \cite{XYLiPRL2007,XHuangAPL2008}.}
It was numerically found that \cite{JZouJAP2009}
the amplitude in the torsional rippling of MWNTs can
be accurately described by a simple sinusoidal shape function,
as confirmed by Figs.~\ref{QWangCarbon2008Fig2undJZouJAP2009Fig4}(d) and (f).
It is noteworthy  that the characteristics of the helical
rippling morphology in twisted MWNTs are different from
those in bent MWNTs, i.e., the so-called Yoshimura
or diamond
buckling pattern. Structurally, torsion-induced rippling
is distributed more periodically and uniformly along the tube
whereas bending-induced rippling is located only in the compressive
side \cite{ArroyoPRL2003,AriasPRL2008}.
Energetically, in twisted MWNTs, 
\color{black}
high strain energy is stored along the ridge regions,
\color{black}
whereas in
bent MWNTs, the strain energy is equally concentrated
\color{black}
both
\color{black}
at the ridges and furrows.


\section{Universal scaling laws under bending and torsion}\label{sec9}

\begin{figure}[ttt]
\centerline{
\includegraphics[width=9.5cm]{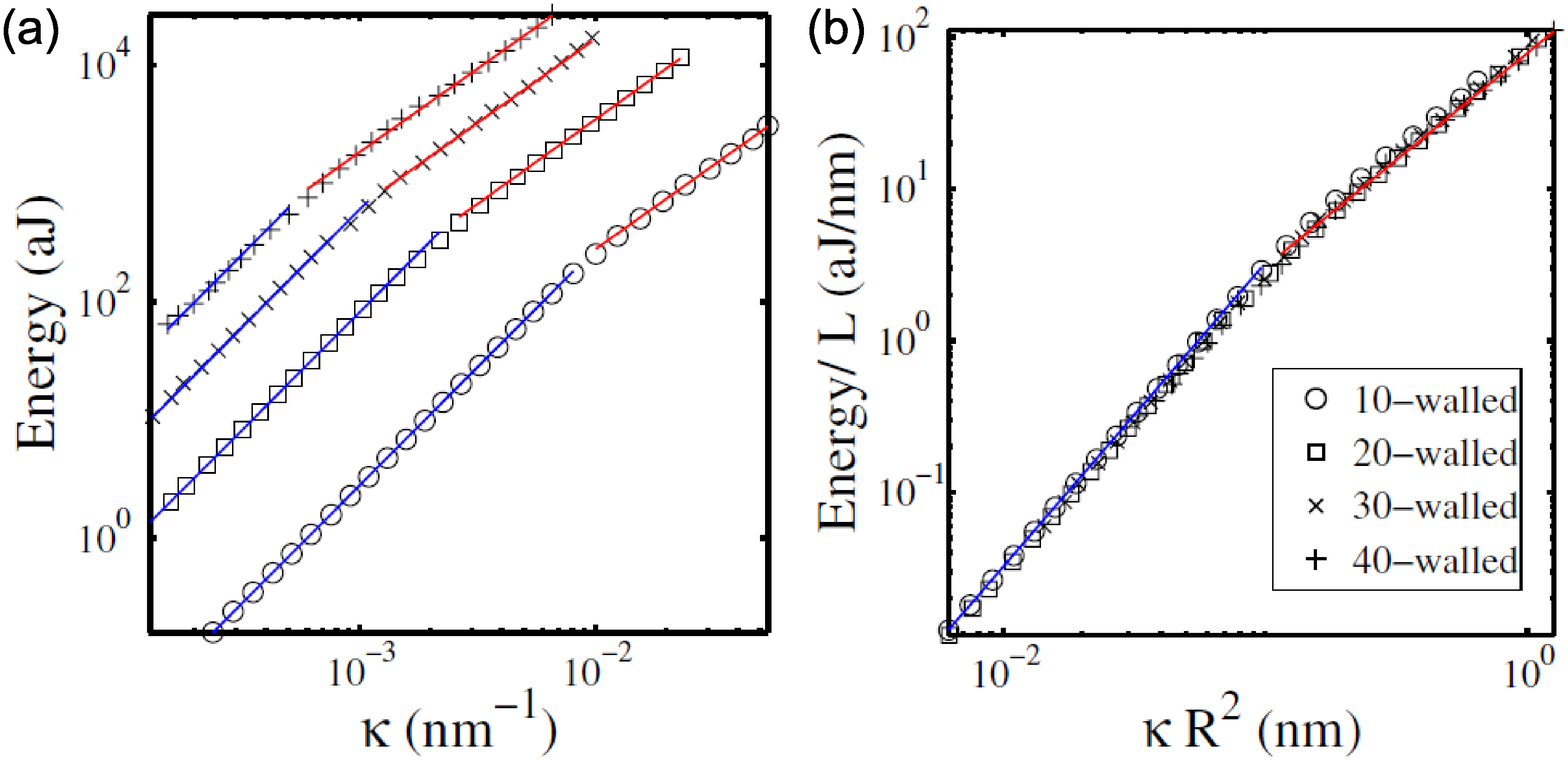}
\includegraphics[width=5cm]{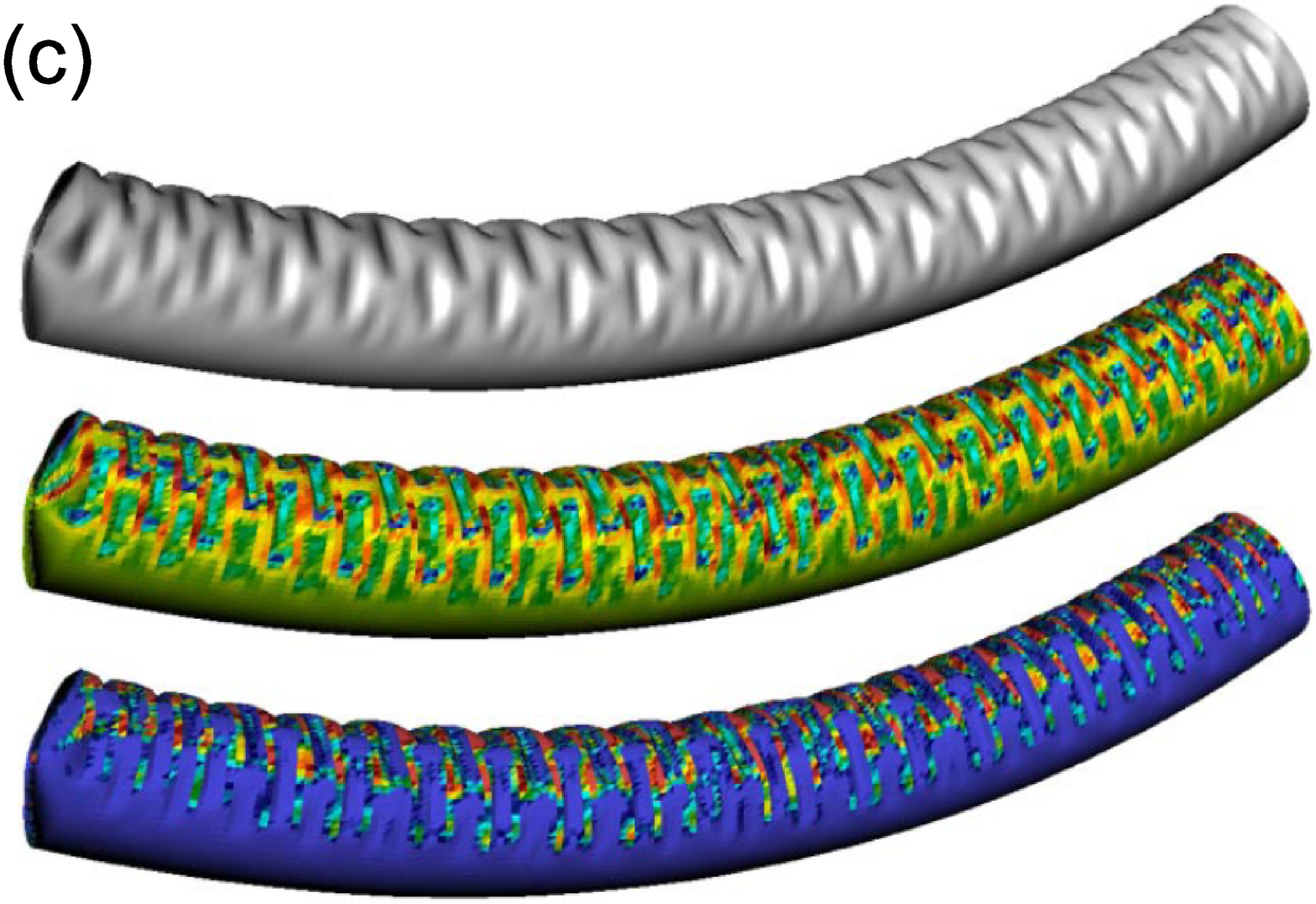}
}
\vspace*{12pt}
\centerline{
\includegraphics[width=9.5cm]{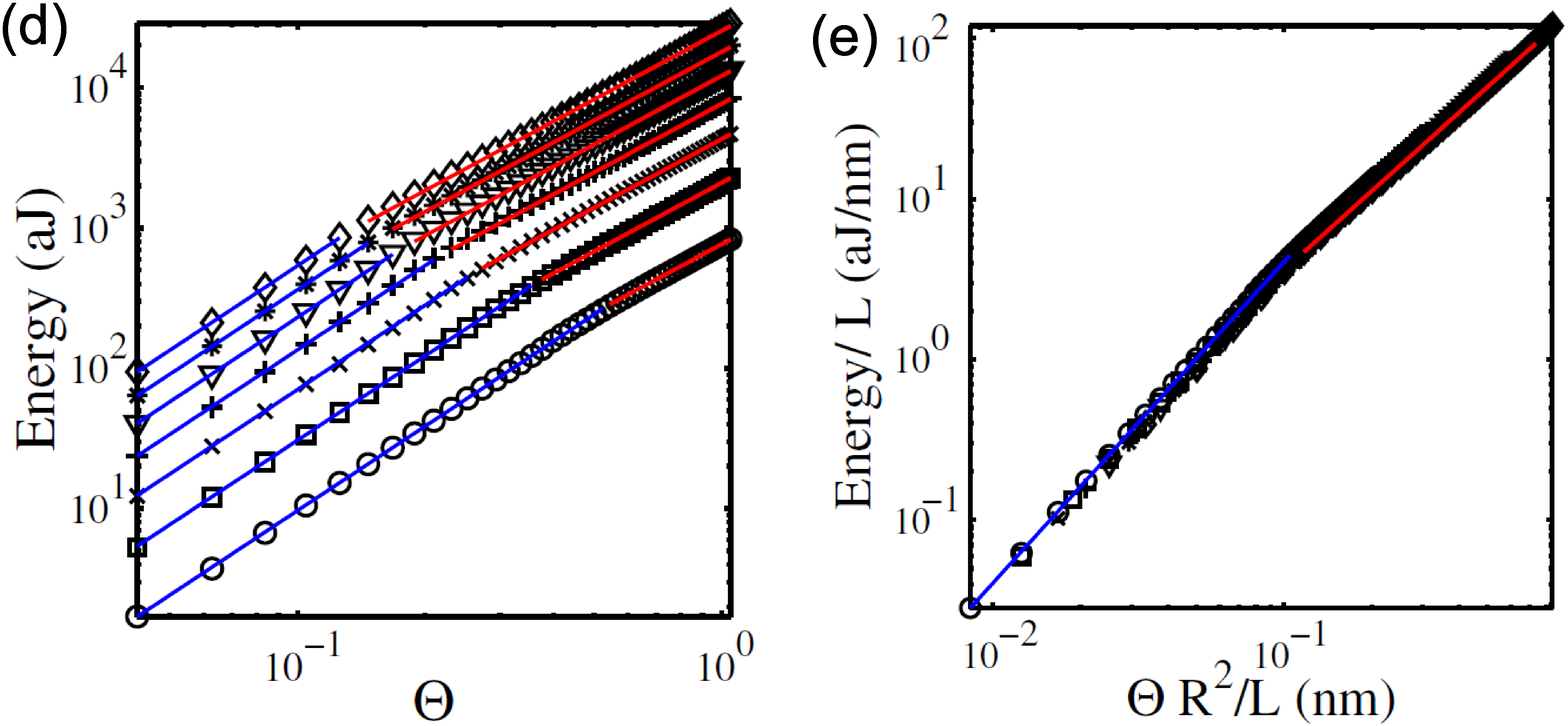}
\includegraphics[width=5cm]{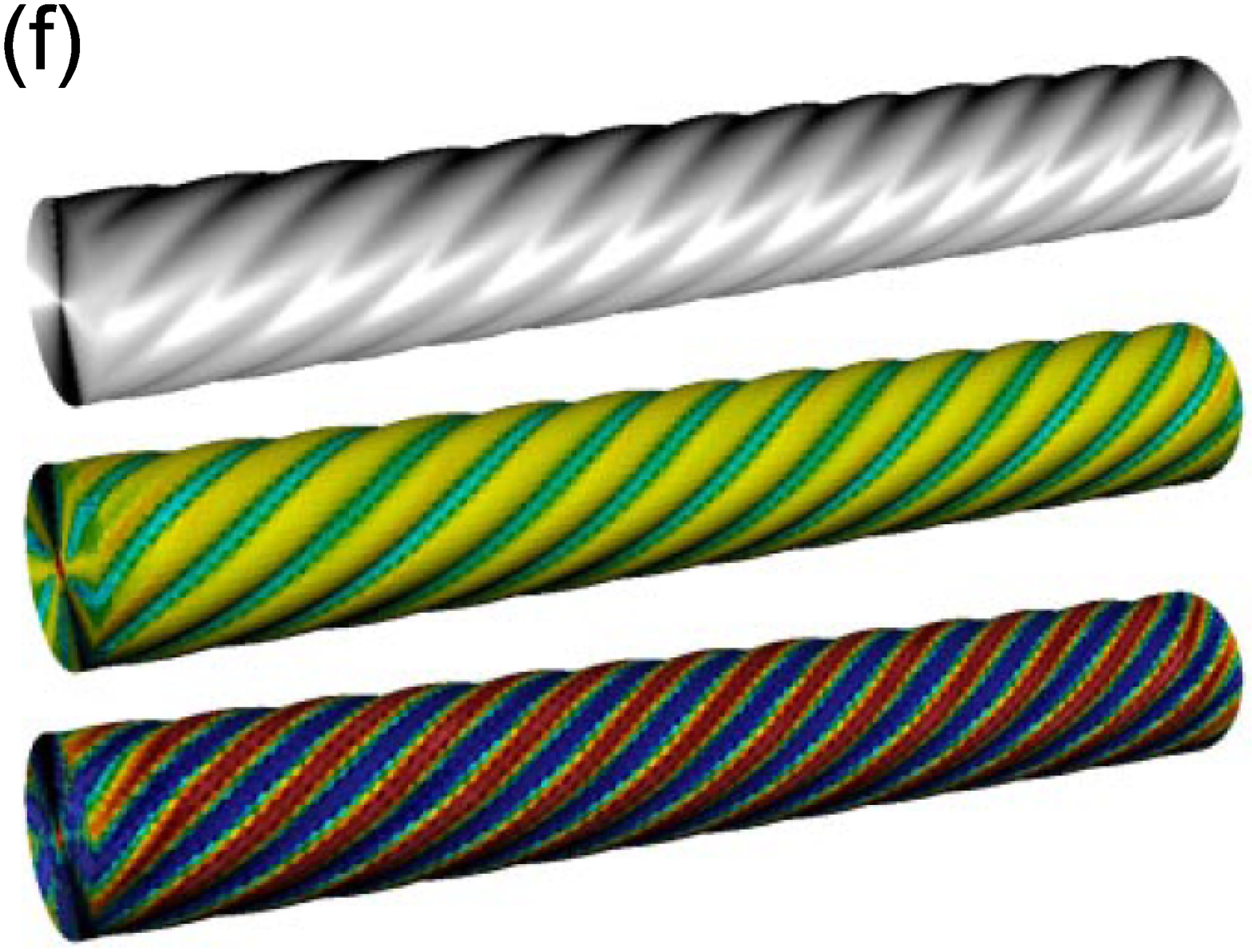}
}
\caption{
(a,d): Strain energy curves as a function of the bending curvature $\kappa$
and the twisting angle $\Theta$.
(b,e): Data collapse upon appropriate rescaling.
The power-law fits with exponents 2 (blue) and $a(<2)$ (red) are shown for illustration.
In all four plots, the number of walls increases stepwise from 10 (circles) to 40 (crosses).
(c): 40-walled nanotube in pure bending.
(f): 35-walled nanotube in torsion.
The latter two panels present the deformed shape (top),
Gaussian curvature map (middle, with green being zero, red being positive, and
blue being negative), and energy density map (bottom, with red being high and
blue being low).
Reprinted from Ref.~\cite{AriasPRL2008}.
}
\label{AriasPRL2008Fig}
\vspace*{60pt}\end{figure}

Interestingly, buckled MWNTs under bending and torsion
were found to obey universal scaling laws that consist of
two distinct power-law regimes in the energy-deflection relation \cite{AriasPRL2008}.
Figure \ref{AriasPRL2008Fig} shows the mechanical response of MWNTs
under bending and torsion;
the strain energy $E$
vs. bending curvature $\kappa$ or twisting angle $\Theta$
is plotted with the increasing stepwise  number of walls from 10 to 40.
All the tested MWNTs exhibit two distinct power-law regimes:
a harmonic deformation regime characterized by the exponent $a=2$
for the relation $E \propto \kappa^a$ or  $E \propto \Theta^a$
(indicated by blue lines in Fig.~\ref{AriasPRL2008Fig})
and an anharmonic, postbuckling regime with exponent $a \sim 1.4$ for bending
and $a\sim 1.6$ for torsion (red lines).
The latter nonlinear response corresponds to the ripples of the graphene walls
 discussed  earlier.

Figures \ref{AriasPRL2008Fig}(b) and (e) show
the data collapse for all the tested nanotubes
upon a universal scaling law.
This law is described by the anharmonic exponent $a$
and the characteristic length scale $\ell_{\rm cr}$.
In twist situations, it is defined by $\ell_{\rm cr} = \Theta_{\rm cr} R^2/L$,
where $L$ and $R$ are the length and outer radius, respectively of the MWNTs considered, and
$\Theta_{\rm cr}$ denotes the critical twisting angle
at which the buckling arises.
In bending cases, $\ell_{\rm cr} = \kappa_{\rm cr} R^2$
with $\kappa_{\rm cr}$ being the critical buckling curvature.
Then, the unified law plotted in red and blue
in Figs. \ref{AriasPRL2008Fig}(b) and (e) is
\begin{equation}
\frac{E(x)}{L} = \propto \left\{
\begin{array}{lc}
(x R)^2 & \mbox{for} \;\; |x R| \le \ell_{\rm cr}, \\ [6pt]
\ell_{\rm cr}^{2-a} |x R|^a & \mbox{for} \;\; |x R| > \ell_{\rm cr},
\end{array}
\right.
\end{equation}
with $x=\Theta R/L$ or $x=\kappa R$.
The actual value of $\ell_{\rm cr}$ was evaluated as $\ell_{\rm cr} \sim 0.1$ nm
for both bending and twisting cases.
It should be emphasized that
since $\ell_{\rm cr}$ has dimension of length,
the unified law is size dependent;
for instance, the thicker the MWNTs,
the smaller will be the obtained $\Theta_{\rm cr}$ or $\kappa_{\rm cr}$.

\color{black}

\section{Challenge and future directions}\label{sec_final}

In this article, I have provided a bird's-eye view on the current state
of knowledge on the buckling properties of carbon nanotubes.
The understanding remains far from complete, 
but new experiments and theoretical work
will no doubt give us a more complete picture
and exciting times for both basic and applied research
in the realm of nanoscale.
Described below are only a few examples of challenging subjects 
that may trigger innovation in the nanotube research community.

\subsection{Buckling effects on heat transport}

Carbon nanotubes demonstrate the excellent thermal conductivity among any known material.
When a carbon nanotube is buckled, however, the localized structural deformation can prohibit 
ballistic heat transport along the nanotube axis \cite{ZHuangJAP2011}.
This results in the decreasing behavior of thermal conductivity of nanotubes
under compressive stress, which is attributed to
the increase in the phonon-phonon scattering rate.
Such the buckling-induced reduction in thermal conductivity
has important implications of heat management of nano-scale electronic devices,
including the dynamic control of thermal transport and energy conversion.
In view of materials sciences, it is also important to make clear 
the effect of buckling on the thermal properties of carbon nanotube composites \cite{FDengAPL2007}.
Despite of great potential, very limited number of studies has been conducted on the issue thus far.

From an academic standpoint, the buckling-induced change in the thermal transport
poses questions on the feasibility of conventional heat conduction theory for
macroscopic solids.
It has been clarified that the low-dimensional nature of carbon nanotubes 
gives rise to various intriguing phenomena:
the well-known examples are
the lattice soliton based energy transfer \cite{CWChangScience2006} 
and 
robust heat transport in deformed nanotubes \cite{CWChangPRL2007}.
These phenomena are beyond the classical way of understanding;
therefore, it is interesting to consider how each class of nanotube buckling 
(compressive, bending, torsional, {\it etc}.) affects the nontrivial heat transport
observed in nanotubes.
The results obtained will shed light on unexplored problems of 
thermal conduction in carbon nanotubes and related materials.

\subsection{Role of defects and imperfections}

Carbon nanotubes obtained by practical synthesis possess
various kinds of defects \cite{ShimaPanPub}
such as missing atoms (called ``vacancy defects"),
carbon rings other than usual hexagonal ones (``Stone-Wales defects"),
or sp$^3$ bonds instead of usual sp$^2$ bonds (``re-hybridization defects").
Hence, understanding the effect of defects on the mechanical properties of nanotubes
is essential in the design of nanotube-based applications.
It has been found that these defects can affect considerably
the mechanical strength and post-elongated morphology of carbon nanotubes.
Comprehensive studies on the influence of defects on their buckling behavior
remained, however, lacking in the literature until recently \cite{CMWangApplMechRev2010}.

MD simulations performed in the past few 
years \cite{HXinCarbon2007,YYZhangJAP2009,YYZhangCarbon2010,KulathungaJPCM2010}
suggested that the presence of defects,
particularly one- or two-atom vacancies,
may cause a significant reduction in the buckling capacity.
An interesting observation is that the degree of reduction
is strongly dependent on the chirality and temperature.
For example, in torsional buckling at low temperature, 
armchair nanotubes are less sensitive to
the presence of defects when compared with their zigzag 
counterparts \cite{YYZhangCarbon2010}.
Still theoretical investigations have been limited to MD simulations;
more accurate and quantitative research based on density-functional methods,
for instance, would be desired.
Experiments on the defect-induced variance in the buckling behavior
are of course to be addressed in future.

Another interesting subject is to employ the presence of defects as regulator of heat conduction
through carbon nanotubes.
With the increase of number of defects, the thermal conductance of nanotubes rapidly decreases. 
The reason for this large reduction is that high-frequency phonons which contribute to thermal transport 
is strongly scattered by the structural defects. 
It has been numerically predicted that \cite{JWangAPL2011}
even a few structural defects in nanotubes can lead to a strong suppression of thermal transport 
by one order of magnitude. 
This result implies that the structural defects can offer an effective method of tuning 
thermal transport of carbon nanotubes; such the tuning of heat transport is advantageous
in the sense that lattice defects can be well controlled during the
growth or by irradiations.

\subsection{Relevance to chemical reaction}

Applications of carbon nanotubes in composite materials often require
an understanding and control of the chemistry and chemical reactivity of
the nanotubes' sidewall.
This is because the carbon-carbon bonding state on the outermost graphitic surfaces
determines where the chemically sensitive reactions take place and how the reactions affect
the physical properties of carbon nanotubes.
For instance, the functionalization and/or chemisorption on the sidewall of nanotubes
enable to increase linking between nanotubes as filler and a surrounding matrix.
Besides, reactivity control of the nanotube sidewall leads to a novel technique for
chemical sensors and drug delivery systems.

An important finding in the context of buckling is that
the chemical reactivity of carbon nanotube is dependent on local surface curvature
of the outermost sidewall.
It has been theoretically demonstrated that \cite{GulserenPRL2001,SParkNanoLett2003}
the reactivity of a nanotube is governed by the local atomic structure of carbon atoms
on which the chemisorbing species or functional group can react and/or a stable bond.
This results imply the control of sidewall reactivity by artificial deformation;
that is, local chemical reactivity of carbon nanotubes can be promoted locally 
by inducing mechanical deformation like buckling.

It should be reminded that 
the basic concept of the curvature-dependent reactivity was initially
proposed in 1993 \cite{HaddonScience1993}; almost twenty years have passed since then.
Nevertheless, research progress on the subject seems 
rather behind \cite{MylvaganamNTN2006,YFZhangCarbon2006}.
Especially, experimental evidence of the proposed reactivity promotion has been scarce,
which may be due to difficulty in precise manipulation of nanotube buckling
and accurate measurement of adsorption capability on the nanotube sidewall.
Considering the importance in view of material science,
breakthrough in the chemistry of the deformation-driven reactivity
and its possible application is strongly expected.

\color{black}

\newpage

\color{black}
\noindent
{\large\bf Appendix: }
\color{black}

\appendix

\section{Who discovered carbon nanotubes first?}\label{secapp}

In 
\color{black}
Appendix,
\color{black}
we take a look at the history of who discovered nanotubes 
and when,\footnote{The relevant literature has been probed with great care
by Boehm in 1997 \cite{BoehmCarbon1997}
and Monthioux and Kuznetsov in 2006 \cite{MonthiouxCarbon2006}.
The first half of this 
\color{black}
Appendix
\color{black}
is based on the two excellent reviews.}
for mainly pedagogical reasons.
The answer seems obvious at a glance, but it is actually quite profound.
The author believes that the argument should be of help for younger readers
who may be learning the facts for the first time.

\subsection{Iijima's nanotube in 1991}

\begin{figure}[ttt]
\centerline{
\includegraphics[width=7.0cm]{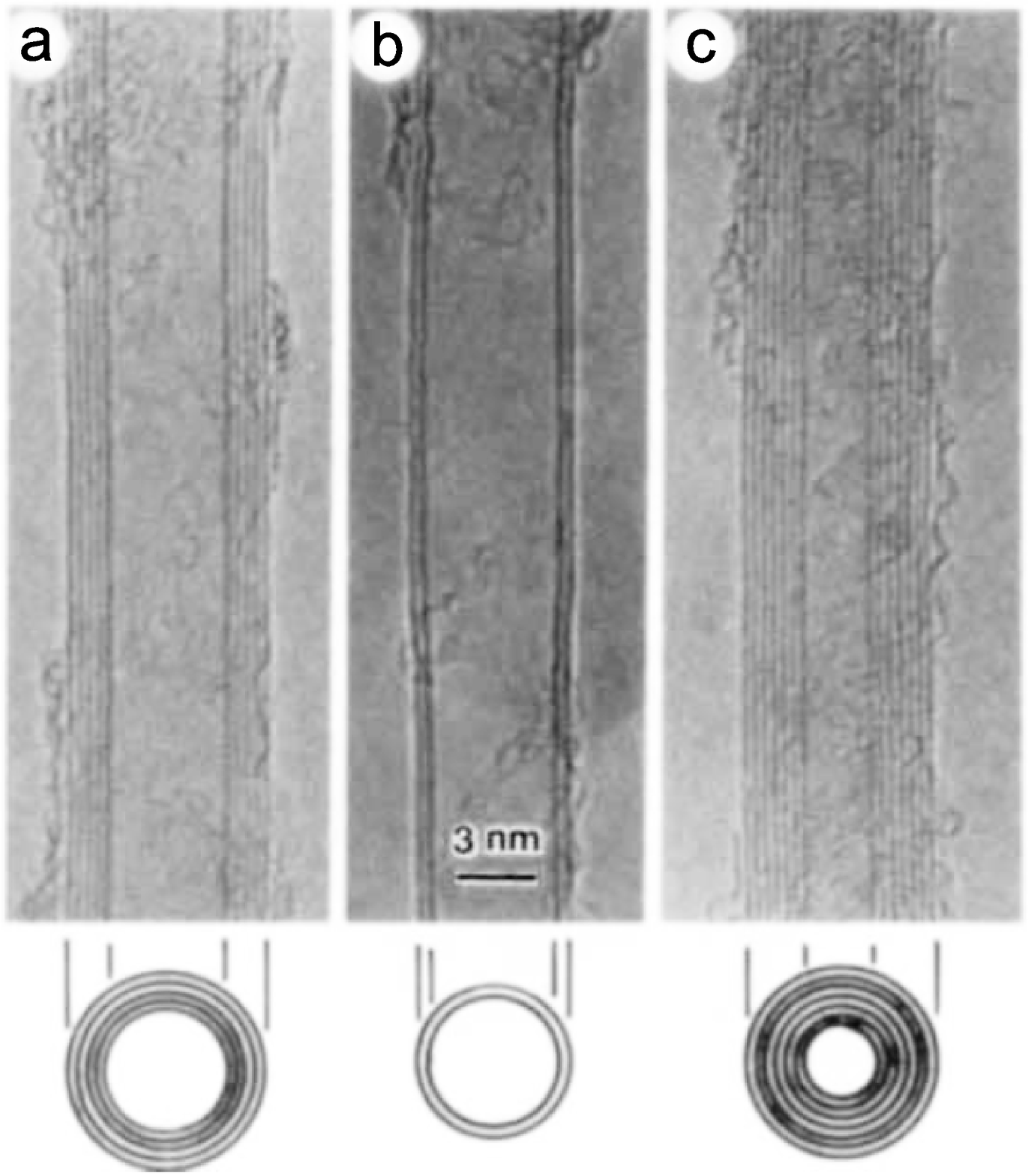}
\includegraphics[width=4.5cm]{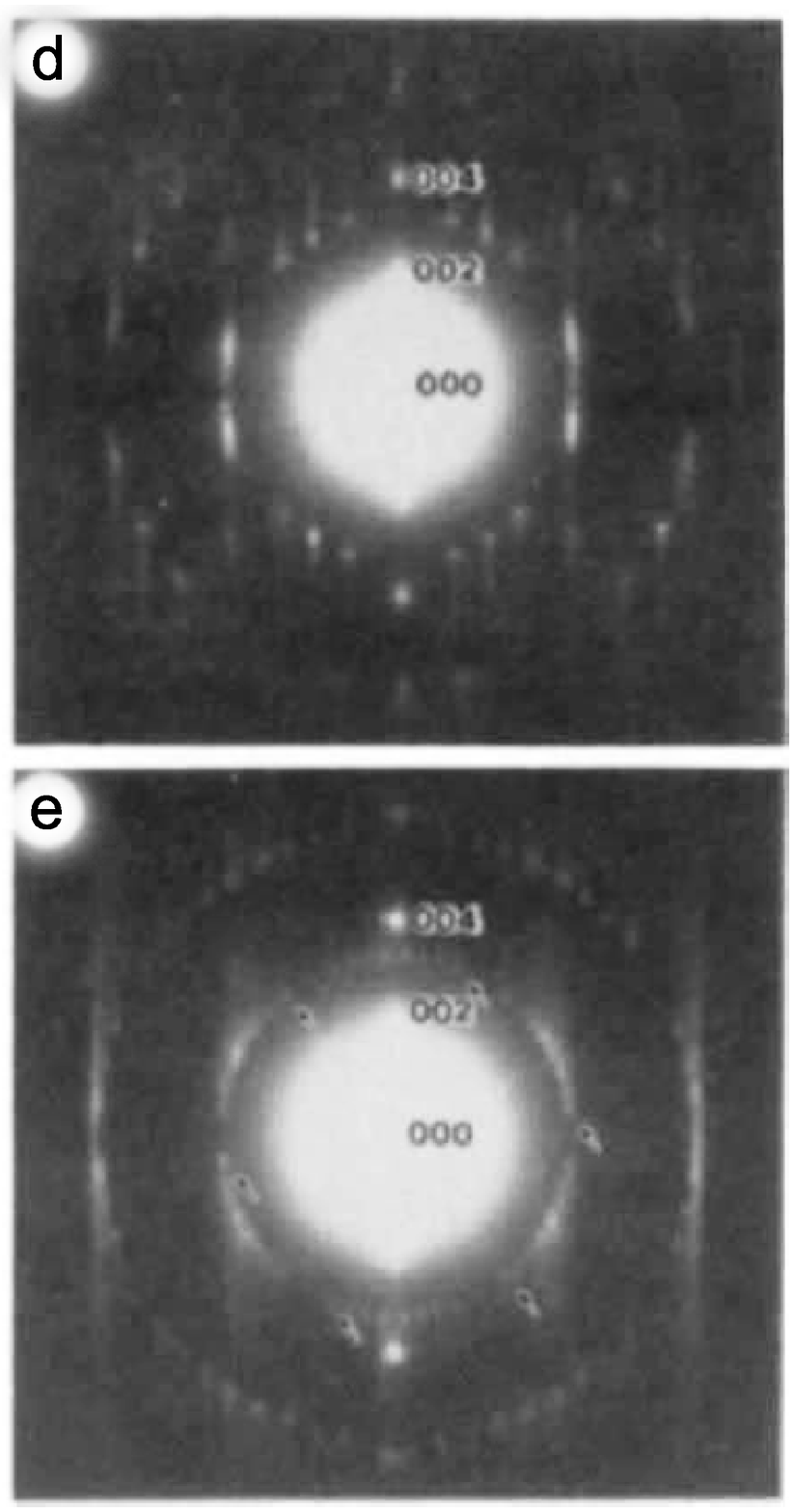}
}
\caption{
TEM image of
nanotubes and sketches of each nanotube's cross sections for
(a) a five-wall nanotube with a  diameter of  6.7 nm,
(b) a double-wall nanotube with a diameter of 5.5 nm, and
(c) a seven-wall nanotube with a diameter of 6.5 nm,
which has the smallest hollow diameter ($\sim$2.2 nm)
among the three specimens.
 Electron diffraction patterns  showing
(d) the superposition of three sets of $\{hk0\}$ spots
taken from a seven-wall nanotube and
(e) the superposition of four sets of $\{hk0\}$ spots
from a nine-wall nanotube.
Reprinted from Ref.~\cite{IijimaNature1991}.
}
\label{IijimaNature1991Fig}
\vspace*{60pt}\end{figure}

As is well known,
most academic and popular literature attributes
the discovery of carbon nanotubes to Sumio Iijima of NEC in 1991.
Without doubt, it is  Iijima's seminal article in 1991 \cite{IijimaNature1991},
entitled ``Helical microtubules of graphitic carbon,"
that triggered the explosion of  carbon nanotube research
still fascinating us today.
Iijima studied the arc evaporation process that efficiently produced
fullerene (C$_{60}$) molecules.
When analyzing a by-product in the arc evaporation,
Iijima found large amounts of MWNTs
mixed with faceted graphitic particles;
see Fig.~\ref{IijimaNature1991Fig} for the image obtained.

From a historical perspective, however,
Iijima's finding is not the first  reported  carbon nanotube.
Careful analysis of the literature shows that
there had been many precedents prior to 1991,
in which  the presence of analogous (and almost identical)
nanostructures including ``carbon tubes" \cite{RadushkevichRuss1952}
and ``hollow carbon fibers" \cite{OberlinJCrystGrowth1976} was uncovered.

\subsection{Carbon nanotubes prior to 1991}

\begin{figure}[ttt]
\centerline{
\includegraphics[width=5.5cm]{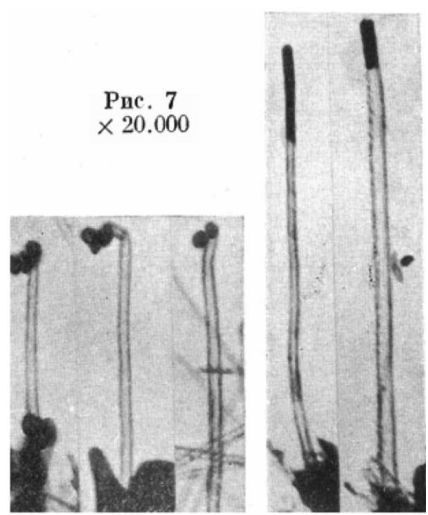}
}
\caption{The earliest TEM images of carbon nanotubes published in
Ref.~\cite{RadushkevichRuss1952}.}
\label{RussiaFig}
\vspace*{60pt}\end{figure}

The first evidence for nanosized carbon tubes is believed to have been published
in 1952 in a Soviet academic journal \cite{RadushkevichRuss1952},
almost 40 years before  Iijima's paper.
The article, written by Radushkevich and Lukyanovich,
demonstrated clear TEM images of tubes (diameter $\sim$ 50 nm) made of carbon.
Figure \ref{RussiaFig} is a reprint of
the image given in Ref.~\cite{RadushkevichRuss1952},
which clearly shows carbon filaments exhibiting a continuous inner
cavity.\footnote{Despite the omission of a scale bar,
the magnification value indicated in the article
allows one to estimate that the diameters of the carbon
tubes imaged are in the range of 50 nm, \textit{i.e}., definitely nanosized.}
However, this discovery was largely unnoticed
as the  article was published in the Russian language.

Subsequently, many other reports followed this Soviet work.
For instance, Oberlin, Endo, and Koyama fabricated in 1976
hollow carbon fibers with nanometer-scale diameters
using a vapor-growth technique \cite{OberlinJCrystGrowth1976}.
In 1979, Abrahamson presented evidence of carbon nanotubes
at a conference paper together with  their characterization
and hypotheses for a  growth mechanism \cite{AbrahamsonCarbon1999}.
In 1981, a group of Soviet scientists produced carbon tubular crystals
and identified them with graphene layers rolled up into cylinders.
Moreover, they speculated that by rolling graphene layers into a cylinder,
many different arrangements of graphene hexagonal nets, such as an armchair-type and a chiral one, are possible.

\subsection{Closing remarks}

In summary, the credit for discovering carbon nanotubes
should go to Radushkevich and Lukyanovich.
It must be emphasized, however, that merely looking at an object with a microscope is completely different
from identifying its structure.
That is, Iijima not only observed
the nanotubes with an electron microscope but also
accurately elucidated its structure from electron diffraction images,
and his work was thus on a different level to preceding studies.

Let us return to Figs.~\ref{IijimaNature1991Fig}(a)--(c).
The photographs shown here are the first electron
micrographs of carbon nanotubes captured by Iijima.
Surprisingly, a multiwalled structure with
an interlayer spacing of less than 1 nm is clearly observed
in this photograph. Even more surprisingly,
he measured the electron diffraction shape [Figs.~\ref{IijimaNature1991Fig}(d) and (e)]
from a single carbon nanotube.
The diffraction strength from a crystal of a light element
(carbon) with a thickness of only several nanometers is
extremely low, and so, this experiment is not easy
to perform even nowadays using the best electron microscopes available.
Iijima must have had first-class microscope operating skills
to make these measurements in 1991, although he did it with ease.
It is also astounding that Iijima identified this substance,
which only appeared as a thin, needle-like crystal,
as a ``tube-shaped substance with a helical structure"
based on its electron diffraction shape.
How many people in the world would have been able to conclude
that it is a nanotube with a helical structure by looking
at the electron diffraction shape?
Now that we already know the structure of carbon nanotubes,
it is not so difficult to read the electron diffraction shape.
However, at that time, much experience
and sharp insight must have been required
by the person observing this electron diffraction shape
for the first time in the world to interpret it correctly
as a tube-shaped substance with a helical structure.

So, who actually discovered carbon nanotubes?
To answer this question, first we need to define the meaning of
``discovery of nanotubes."
It is probably correct to state that Radushkevich and Lukyanovich
were the first to take photographs of the overall picture of
the cylindrical tube-shaped nanocarbon substance
(although photographs taken by other researchers could still be discovered in future).
However, Iijima was the first to reveal that the microscopic structure
unique to nanotubes was multiwalled and helical,
by using ``the eyes" of electron diffraction.
The rest is up to the philosophy and preference of
the person who is judging.
Whatever the judgment, we should never forget that the knowledge
and benefits that we now enjoy in nanotube research are
the result of the hard work and achievements of our predecessors.


\vspace*{60pt}


\section*{Acknowledgments}

I cordially thank  Prof.~Motohiro~Sato at Hokkaido University,
Prof.~Marino~Arroyo at Universitat Polit\`ecnita de Catalunya, and
Dr.~Susanta~Ghosh at University of Michigan for illuminating discussions.
I also deeply thank Prof.~Emeritus~Eiji~Osawa at Toyohashi University of Technology,
Prof.~Jun~Onoe at Tokyo Institute of Technology,
and Prof.~Hideo Yoshioka at Nara Women's University,
as communications with them have sparked my interest in the study of
nanocarbon materials.
\color{black}
Helpful suggestions from three anonymous reviewers are acknowledged.
\color{black}
This work was supported by MEXT,
the Inamori Foundation, and the Suhara Memorial Foundation.
Lastly, the most gratitude is owed to Prof. Kosuke Yakubo in Hokkaido University,
for his generous support and encouragement during the completion of the project.
\vspace*{24pt}


\bibliographystyle{mdpi}
\makeatletter
\renewcommand\@biblabel[1]{#1.}
\makeatother



\end{document}